\documentclass[english]{article}
\usepackage[T1]{fontenc}
\usepackage[latin9]{inputenc}
\usepackage{geometry}
\usepackage[active]{srcltx}
\usepackage{float}
\usepackage{amsmath}
\usepackage{amssymb}
\usepackage{graphicx}

\makeatletter

\providecommand{\tabularnewline}{\\}
\let\@fnsymbol\@arabic

\makeatother

\usepackage{babel}

\title{\vspace{-0.5cm}Modeling the deep drawing of a 3D woven fabric with a second gradient model}

\author{	
	Gabriele Barbagallo\footnote{Gabriele Barbagallo, gabriele.barbagallo@insa-lyon.fr, LaMCoS-CNRS \& LGCIE, INSA-Lyon, Universitité de Lyon, 20 avenue Albert Einstein, 69621, Villeurbanne cedex, France} \ and 
	Angela Madeo\footnote{Angela Madeo, corresponding author, angela.madeo@insa-lyon.fr, LGCIE, INSA-Lyon, Université de Lyon, 20 avenue	Albert Einstein, 69621, Villeurbanne cedex, France}  \ and 
	Fabrice Morestin\footnote{Fabrice Morestin, fabrice.morestin@insa-lyon.fr, LaMCoS-CNRS, INSA-Lyon, Universitité de Lyon, 20 avenue Albert Einstein, 69621, Villeurbanne cedex, France}  \ and 
	Philippe Boisse\footnote{Philippe Boisse, philippe.boisse@insa-lyon.fr, LaMCoS-CNRS, INSA-Lyon, Universitité de Lyon, 20 avenue Albert Einstein, 69621, Villeurbanne cedex, France} 
}

\begin{document}

	\maketitle

	\addtocounter{footnote}{4} 
\vspace{-0.85cm}	
\begin{abstract}
Experimental testing on dry woven fabrics exhibits a complex set of evidences that are difficult to be completely described using classical continuum models. The aim of this paper is to show how the introduction of energy terms related to the micro-deformation mechanisms of the fabric, in particular to the bending stiffness of the yarns, helps in the modeling of the mechanical behavior of this kind of materials. To this aim, a second gradient, hyperelastic, initially orthotropic continuum theory is proposed to model fibrous composite interlocks at finite strains. In particular, the present work explores the relationship between the onset of wrinkling appearing during the simulation of the deep drawing of a woven fabric and the use of a second gradient model. It is shown that the introduction of second gradient terms accounting for the description of in-plane and out-of-plane bending rigidities, decreases the onset of wrinkles during the simulation of deep-drawing.

In this work, a quadratic energy, roughly proportional to the square of the curvature of the fibers, is presented and implemented in the simulations. This simple constitutive assumption allows to clearly show the effects of the second gradient energy on both the wrinkling description and the numerical stability of the model. The results obtained in second gradient simulations are descriptive of the experimental evidence of deep drawing whose description is targeted in this work. The present paper provides additional evidence of the fact that first gradient continuum theories alone cannot be considered fully descriptive of the behavior of dry woven composite reinforcements. On the other hand, the proposed second gradient model for fibrous composite reinforcements opens the way both to the more accurate simulation of complex forming processes and to the possibility of controlling the onset of wrinkles.
\end{abstract}
\vspace{-0.5cm}		
	\tableofcontents
	
	\addtocontents{toc}{\vspace{-0.3 cm}}

	\newpage
	
\section{Introduction}

Composite materials may possess very good characteristics due to the fact that they are obtained assembling two or more constituent materials. This kind of architectured materials can be optimized to obtain excellent resulting properties such as high strength, lightness, cost-effectiveness, thermal and electrical conductivity, insulation and many others. 

In particular, the fibrous composite materials are a class of composites that is manufactured using strong fibers aligned in unidirectional sheets, non-crimped fabrics or woven fabrics (bidirectional), impregnated with a thermoset or thermoplastic resin. Woven composite materials are the most widespread choice in the case of mechanical reinforcement, due to their great formability and the subsequent possibility of designing rather complex mechanical pieces. The forming processes, such as the Resin Transfer Moulding, (RTM), have received a great deal of attention in the literature (see e.g. \cite{parnas2000liquid,potter1999early,rudd1997liquid}). The most current forming processes consists basically of two stages:
\begin{itemize}
	\item the dry woven fabric is preformed to obtain the desired geometry for the final part 
	\item a thermoset resin is injected into the woven fabrics filling the pores of the fibrous reinforcement.
\end{itemize}
The process used in the forming stage is thus followed by the injection and curing of a resin in  the woven fabrics, after which the finished material, union of the reinforcement and the matrix, is obtained. The quality of the obtained piece is greatly influenced by several factors, such as the characteristics of the preformed woven fabric and, in particular, its permeability, the characteristic of the resin and the temperature at which the injection process takes place. In the literature, it is possible to find a great number of articles studying in detail the injection processes and the characteristic of the resins (see \cite{parnas2000liquid,potter1999early,rudd1997liquid}) and also the preforming processes of thin woven reinforcements (see \cite{boisse2011simulation,boisse2001meso,boisse1995experimental,boisse2008different,boisse2005mesoscopic,deluycker2009simulation,deluycker2010experimental,gatouillat2013meso,hamila2007meso,wang2015experimental,zouari2006woven}). Nonetheless, few research work concern 3D composite reinforcement forming simulation \cite{charmetant2012hyperelastic,mathieu2014locking,pazmino2015numerical}. The focus of this paper will be on the first stage of the forming process, namely the preforming of 3D dry reinforcements. The understanding of this step is very important to determine if the preforming process is even possible. Indeed, the woven fabrics can withstand only a certain amount of shear deformation between the fibers without dissociating and an accurate modeling thus becomes crucial for optimal design.

\begin{figure}[H]
	\begin{centering}	
		\includegraphics[height=7cm]{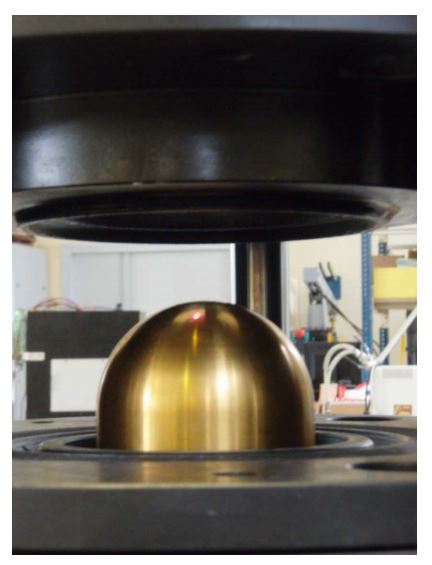} 
		\includegraphics[height=7cm]{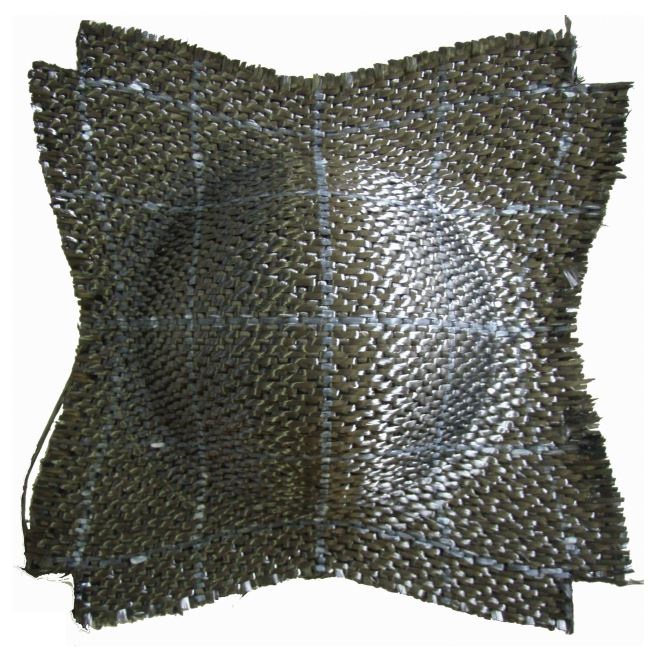}
		\par\end{centering}
	\caption{Experimental setup  and deformation for a deep-drawing preforming with a hemispherical punch \cite{charmetant2012hyperelastic}.\label{deepdrawing}}	
\end{figure}

The process of preforming can become fairly intricate when the geometries are complex (e.g. double curved geometries) and the prediction of the entirety of the properties of the deformed fabrics is, therefore, challenging. Several experimental devices have been set up to investigate the deformation modes and the possible occurrence of defects during forming of textile reinforcements \cite{cao2008characterization,lee2008bias}. Among them, the hemispherical punch and die systems (Fig.\;\ref{deepdrawing}) were especially studied because of their simple shape, double curvature and large shear angle variations between the yarns in the final state.

Reliable models for the preforming process should include information about the fiber directions and densities in the deformed state, so aiding the simulation of the resin injection and of the structural behavior of the final composite part. Indeed, the permeability of the 3D interlock fabric is strongly influenced by some deformation states that can alter or even close the interstices in the micro-structure affecting profoundly the resulting material properties. Furthermore, the direction and positioning of the yarns, that are determined solely by the preforming process, have a predominant role in the resulting mechanical properties of the composite structure (stiffness, damage and fracture, etc.).

Different approaches have been proposed  to model the raw fibrous composite materials that can be found in the literature (see \cite{boisse2001meso,boisse2008different,boisse2005mesoscopic,deluycker2009simulation,deluycker2010experimental,gatouillat2013meso,hamila2007meso}). The most widespread approach to the simulation of fibrous composite materials is, nowadays, the finite element model that needs the determination of specific constitutive laws, to be able to describe the complex experimental evidences shown by woven composites. The interest of using second gradient theories for the more realistic description of the mechanical behavior of fibrous composite reinforcements has already been established in previous contributions (see \cite{barbagallo2016bias,ferretti2014modeling,madeo2016continuum,madeo2015thick}). 

In the present paper, additional evidence is provided regarding the fact that neglecting the bending rigidities of the yarns in the modeling phase can produce inaccurate results of the simulation of 3D woven fabrics (thickness $\sim$1 cm) during the modeling phase. In order to support this statement, a 3D FEM is implemented and a rather simple constitutive form of the strain energy density is introduced, accounting for:
\begin{enumerate}
	\item initial orthotropy,
	\item geometric non-linearities,
	\item in-plane and out-of-plane bending of the yarns (through the introduction of suitable second gradient terms).
\end{enumerate}
This second gradient model is implemented in COMSOL$^{\circledR}$ looking for solutions that are continuous, as it is usual in FEM, but that also grant continuity of the first derivatives of the displacement field.

In this way, the following desirable results are obtained:
\begin{itemize}
	\item the solution is in agreement with the observed experimental shapes (Fig. \ref{deepdrawing}),
	\item the second gradient energy has a beneficial effect on the mesh-dependency of the solution,
	\item the presence of suitable second gradient terms which are descriptive of the yarns' bending allows to control the onset and evolution of wrinkles during the deep-drawing process. More particularly, if the second gradient parameter (viz the bending stiffness of the yarns) is sufficiently high, no wrinkling is observed during the simulation. This result is in agreement with the common observations of experimental results.
	
	The results presented in this paper should be used as a guide towards the throughout implementation of FEM codes including second gradient constitutive laws for the complete modeling of the mechanical behavior of fibrous composite reinforcements during their forming process.
\end{itemize}

\section{Second gradient 3D modeling of the deep drawing}

The description of woven composites' mechanical behavior  demands important efforts. Through the analysis of the deformation patterns during experimental testing, it is easy to notice that the condition of material continuity is not always strictly fulfilled, due for example to some relative slippage of warp and weft. However, if the amount of slippage between the fibers is low, a continuous model can still be used (see for example  \cite{charmetant2012hyperelastic,charmetant2011hyperelastic}).  This is the approach adopted in this paper, but it must be noted that the possibility of modeling each fiber as a single detached element still exists, even if it is of difficult applicability for big mechanical pieces \cite{durville2005numerical}. Continuum models with ``fictive'' elongations have been also introduced to account for a certain amount of slipping while remaining in a continuum framework \cite{madeo2016continuum}.

The composite reinforcement  described in the present paper is a 3D interlock fabric. It is 1 cm thick and it is composed of  two yarn directions that are woven together and through the thickness. This 3D fabric will be analyzed with 3D Finite Elements.

The most crucial part of the continuous modeling is the definition of proper constitutive relationships that realistically reflect the mechanical properties of the analyzed material. Traditionally, the energy used in the simulation of continuous media comprises only deformations defined as first derivative of the displacement, so giving rise to so called first gradient theories. However, in various papers dealing with woven composites, it is shown how the addition of energies related to the local stiffness of the yarns is useful,  if not necessary, to describe the macroscopic deformation behavior of the interlocks (e.g. \cite{barbagallo2016bias,charmetant2012hyperelastic,ferretti2014modeling,madeo2015thick,madeo2016continuum}). 

Considering the specific case of deep-drawing preforming, one of the phenomena which is most difficult to control with a first gradient energy is the onset of wrinkling in the deformed 3D fabric. In first gradient simulations, it is possible to observe the presence of wrinkles in the deformed configurations when a certain amount of in-plane shear stiffness is present (see for example \cite{boisse2006importance}). Nevertheless, the number and amplitude of such wrinkles is a mesh-dependent phenomenon and such wrinkling is not descriptive of the experimental results. This result will be obtained again in the present paper for a traditional first gradient finite element model with linear shape functions (subsection \ref{sec:Mesh-1}).

In the remainder of the paper, it will be shown how the necessary description of the local bending  effects can be performed via generalized continuum theories, such as higher order gradient or constrained micromorphic theories. More precisely, a second gradient energy, approximately proportional to the square of the yarns' curvature, is introduced and the obtained results are used to model the deep-drawing of woven fabrics. 

The generalized continuum theories are still perceived as pure theoretical abstractions, even if their background was laid down in detail since the historic works of Piola \cite{piola1846memoria}, Cosserat \cite{cosserat1909theorie}, Midlin \cite{mindlin1964micro}, Toupin \cite{toupin1964theories}, Eringen \cite{eringen1999microcontinuum}, Green and Rivlin \cite{green1964multipolar} and Germain \cite{germain1973method}. With the hope of a more widespread consciousness of the potential of generalized continuum theories, the authors attempt in this paper to show how some microstructure-related effects in microscopically heterogeneous mechanical systems can be still modeled by means of continuum theories.

\subsection{Second gradient modeling}

In the remainder of the paper, the following notations will be used:
\begin{itemize}
	\item $\ensuremath{B_{L}\subset\mathbb{R}^{3}}$  is the Lagrangian or reference configuration of the considered continuum and each point $\mathbf{X}\in B_{L}$ is called a material point,
	\item $\boldsymbol{\chi}(\mathbf{X},t):B_{L}\times[0,T]\rightarrow\mathbb{R}^{3}$ is a suitably regular kinematic field which associates to any material point $\mathbf{X}$ its current position $\ensuremath{\mathbf{x}}$ at the time t,
	 \item  $\ensuremath{\mathbf{u}(\mathbf{X},t):=\boldsymbol{\chi}(\mathbf{X},t)-\mathbf{X}}$ is the displacement field at the time t given by the difference between the current and the reference position of each material point $\mathbf{X}$,
	 \item $\ensuremath{B_{E}(t)}$ is the current shape of the body at any instant t, usually called  the Eulerian configuration, given by the image of the function $\boldsymbol{\chi}(B_{L},t)$ ,
	 \item the tensor $\ensuremath{\mathbf{F}:=\nabla\boldsymbol{\chi}}$ is the gradient of the map $\boldsymbol{\chi}$ with respect to the reference position $\mathbf{X}$ ,
	 \item $\mathbf{C}:=\mathbf{F}^{T}\cdot\mathbf{F}$  is the Right Cauchy-Green deformation tensor\footnote{A central dot indicates simple contraction between tensors of order greater than zero. For example if $\mathbf{A}$ and $\mathbf{B}$ are second order tensors of components $A_{ij}$ and 	$B_{jh}$ respectively, then $(\mathbf{A}\cdot\mathbf{B})_{ih}:=A_{ij}B_{jh}$, 	where Einstein notation of sum over repeated indexes is used.},
	 \item the third order tensor field $\nabla\mathbf{C}$ is the gradient of the Cauchy-Green deformation tensor.
\end{itemize}
Compared to the first gradient models, an additional third order tensor field $\nabla\mathbf{C}$ is introduced in the xonstitutive expression of the strain energy density. This tensor field makes it possible to decribe effects related to the macro-inhomogeneity due to micro-deformations in the micro-structure of the continuum, such as the curvature of the yarns. Since second gradient theories can be readily obtained as limiting cases of micromorphic ones, it is possible to derive the second gradient contact actions in terms of the micromorphic ones following the procedure used in \cite{bleustein1967note}. Some of the possible types of constraints, that could be included in such a micromorphic model which, for example, impose inextensibility of yarns so giving rise to so-called micropolar continua, are presented in \cite{altenbach2010acceleration,eremeyev2005acceleration,eremeyev2013foundations,eringen1999microcontinuum,pietraszkiewicz2009vectorially}. 

A hyperelastic, initially orthotropic, second gradient model can be applied to the case of thin fibrous composite reinforcements at finite strains. For the strain energy density $W\left(\mathbf{C},\ \nabla\mathbf{C}\right)$, which will be used to simulate the mechanical behavior of the fibrous materials in the finite strain regime, it is assumed a decomposition such as:
\begin{equation}
W\left(\mathbf{C},\ \nabla \mathbf{C}\right)=W_{I}(\mathbf{C})+W_{II}(\nabla\mathbf{C}),
\end{equation}
where $\ensuremath{W_{I}}$ is the first gradient strain energy and $\ensuremath{W_{II}}$ is the second gradient one. The specific constitutive forms of the first and second gradient strain energy densities used to model fibrous composite reinforcements are explicitly presented in the following subsections.

\subsection{Hyperelastic initially orthotropic first gradient strain energy density}

Even at finite strains, well-known expressions for isotropic strain energies descriptive of the behavior of isotropic materials are available in the literature (see e.g. \cite{ogden1984nonlinear,steigmann2002invariants}). Quite the opposite happens in the case of orthotropic materials, for which suitable specific strain energies, well descriptive of real material behaviors are more difficult to be found. Some results are provided in  \cite{itskov2004class}, where some polyconvex energies are proposed to describe the deformation of rubbers in uniaxial tests. Explicit anisotropic hyperelastic potentials for soft biological tissues are also proposed in \cite{holzapfel2000new} and reconsidered in \cite{balzani2006polyconvex,schroder2005variational}, in which their polyconvex approximations are derived. Other examples of polyconvex energies for anisotropic solids are given in \cite{steigmann2003frame}. 

Notwithstanding the research efforts devoted to the study of polyconvexity, which certainly introduce rigorous theoretical frameworks for the study of the mechanical behaviors of hyperelastic materials, the use of such polyconvex models is often limited due to the difficult attribution of a sensible physical meaning to the wealth of constitutive parameters which are introduced. The approach adopted in this paper is the Ockham's razor approach, introducing the minimum possible number of physically sensible constitutive parameters which are needed to describe the targeted phenomena.

In the literature, reliable constitutive  models for the description of the mechanical behavior of fibrous composite reinforcements at finite strains can be found in \cite{aimene2010hyperelastic,charmetant2012hyperelastic,charmetant2011hyperelastic}. Moreover, the mechanical behavior of composite preforms with rigid organic matrix (see e.g. \cite{mikdam2009effective,mikdam2010statistical,oshmyan2006principles}) is quite different from the behavior of the sole fibrous reinforcements (see e.g. \cite{charmetant2012hyperelastic}) rendering the mechanical characterization of such materials a major scientific and technological issue. 

\medskip

In this work, the directions $\ensuremath{\mathbf{D}_{1}}$ and $\ensuremath{\mathbf{D}_{2}}$ denote the unit vectors in the directions of the warp and weft yarns in the reference configuration and the direction $\ensuremath{\mathbf{D}_{3}}=\mathbf{D}_{1}\times\mathbf{D}_{2}$ denotes the unit normal to the plane containing the two sets of fibers. It is possible to fully describe a first gradient orthotropic energy with an expression of the type (see e.g.  \cite{raoult2009symmetry}):
\begin{equation}
W_{I}(\mathbf{C})=W_{I}(i_{11},i_{22},i_{33},i_{12},i_{13},i_{23}),
\end{equation}
where $i_{ii} =\mathbf{D}_{i}\cdot\mathbf{C}\cdot\mathbf{D}_{i},\ i=\{1,2,3\}$ represents the elongation strain in the direction $\ensuremath{\mathbf{D}_{i}}$ and $i_{ij}=\mathbf{D}_{i}\cdot\mathbf{C}\cdot\mathbf{D}_{j}$ represents the shear strain (angle variation) between the directions $\ensuremath{\mathbf{D}_{i}}$ and $\ensuremath{\mathbf{D}_{j}}$ with $i,\,j\,\in\{1,2,3\}$ and $i\neq j$.

It is possible to develop complex non-linear energies that capture in all details the mechanical non-linearities observable in the experimental testing, as done in \cite{aimene2010hyperelastic,charmetant2012hyperelastic,charmetant2011hyperelastic}, but, as already pointed out, this is not one of the aims of the present paper. Instead, using only a simple quadratic first gradient energy, it is possible to thoroughly analyze the influence of both meshing and additional second gradient terms on the performed numerical simulations. Thus, the chosen constitutive expression for the first gradient energy is:
\begin{align} W_{I}(\mathbf{C})=\frac{1}{2}K_{11}(\sqrt{i_{11}}-1)^{2}+\frac{1}{2}K_{22}(\sqrt{i_{22}}-1)^{2}+\frac{1}{2}K_{33}(\sqrt{i_{33}}-1)^{2}+\frac{1}{2}K_{12}i_{12}^{2}+\frac{1}{2}K_{13}i_{13}^{2}+\frac{1}{2}K_{23}i_{23}^{2}, \label{First}\end{align}
where  $K_{ii}$ are the extensional stiffnesses in the direction of the yarns as well as in the orthogonal direction, while $K_{ij}$ with $i\neq j$ are the in-plane and out-of plane shear stiffnesses. The numerical values of the material parameters were chosen to define a material in which the extensional stiffness is much higher than the shear stiffness and the shear behavior in the plane of the fibers is stiffer than the out of plane. Moreover, the extensional stiffness in the orthogonal plane is much lower than the in-plane ones, due to the fact that no yarns are effectively present in the thickness of the interlock. Even if more refined hyperelastic laws can be certainly be introduced in the spirit of \cite{aimene2010hyperelastic,charmetant2012hyperelastic,charmetant2011hyperelastic}, the proposed expression for the first gradient energy density is representative of the main macroscopic deformation modes of fibrous interlocks. The parameters chosen are the ones shown in Tab.\;\ref{table: Par}.

\begin{table}[H]
	\begin{centering}
		\begin{tabular}{|c|c|c|c|c|c|c|}
			\hline 
			$K_{11}$ & $K_{22}$ & $K_{33}$ & $K_{12}$& $K_{13}$& $K_{23}$\tabularnewline
			\hline 
			\hline 
			5 MPa & 5 MPa &  0.5 MPa & 50 kPa & 0.5 kPa & 0.5 kPa \tabularnewline
			\hline 
		\end{tabular}
		\par\end{centering}
	
	\caption{Parameters of the first gradient energy.\label{table: Par}}
\end{table}

\subsection{Hyperelastic orthotropic second gradient strain energy density}

Considering linear elastic isotropic second gradient media, it is possible to find constitutive laws that are able to describe a very wide set of behaviors (see for example \cite{dellisola2009generalized}). In the case of the woven fabrics, the bending stiffness of the yarns is the main micro-structure-related deformation mechanism which takes place at the mesoscopic level and, therefore, it is the only one that will be considered in this paper. The modeling of the bending stiffness of the yarns is decisive for the description of some specific phenomena, such as shear transition layers in 2D experimental tests and wrinkling during the deep-drawing of dry woven fabrics. A second gradient theory is potentially able to account for other effects related to the derivatives of the elongations but, in this work, they will be disregarded. The second gradient energy considered is, thus, function only of the derivatives of the invariants $i_{ij}$ ($i\neq j$), that can be used to define rough descriptors of the curvatures of the two sets of yarns of the fabric. 

As a matter of fact, it can be inferred (see also \cite{barbagallo2016bias,dellisola2015two,ferretti2014modeling,madeo2016continuum,madeo2015thick}) that, given the family of yarns initially oriented in the direction $\mathbf{D}_{1}$, the quantity $i_{12,1}$ is a measure of their in-plane bending\footnote{Here and in the sequel the term $(\cdot)_{,i}$ denotes the partial derivative of the quantity $(\cdot)$ with respect to the space coordinates $\xi_{i}$ of a reference frame oriented within the directions $\mathbf{D}_{i}$.}. Analogously $i_{12,2}$ is a measure of the in-plane bending of the family of yarns initially oriented in the direction $\mathbf{D}_{2}$. The quantities $i_{13,1}$ and $i_{23,2}$ are descriptors of the out-of-plane bending of the yarns initially oriented in the $\mathbf{D}_{1}$ and $\mathbf{D}_{2}$ directions, respectively. Since no material fibers are present in the thickness of the considered interlocks, quantities related to their bending ($i_{13,3}$ and $i_{23,3}$) are not likely to play a role in the deformation of such materials. In the light of these remarks, the following constitutive form is introduced for the second gradient strain energy density:
\begin{equation}
W_{II}(\nabla \mathbf{C})=\frac{1}{2}\,\alpha_{1}\, i_{12,1}^{2} +\frac{1}{2}\,\alpha_{2}\, i_{12,2}^{2}+ \frac{1}{2}\,\beta_{1}\,i_{13,1}^{2}+\frac{1}{2}\,\beta_{2}\, i_{23,2}^{2},
\end{equation}
where with $\alpha_{1}$, $\alpha_{2}$ and $\beta_{1}$, $\beta_{2}$ are the in-plane and out-of-plane bending stiffnesses of the two family of yarns, respectively. For unbalanced fabrics, i.e. fabrics whose warp and weft yarns do not have the same characteristics, it is likely that $\alpha_{1}\neq\alpha_{2}$ and $\beta_{1}\neq\beta_{2}$ (see also \cite{barbagallo2016bias,madeo2016continuum}). The object of this paper are interlocks which are balanced and, hence, it is assumed that $\alpha_{1}=\alpha_{2}=\alpha$ and $\beta_{1}=\beta_{2}=\beta$. Moreover, it is possible that the two families of yarns have different bending stiffnesses in-plane and out of plane. Nevertheless, such difference can be rather small, so in this paper it will also be set $\alpha=\beta$. The chosen second gradient energy thus takes the particular form:
\begin{equation}
W_{II}(\nabla \mathbf{C})=\frac{1}{2}\alpha \left( i_{12,1}^{2} + i_{12,2}^{2}+ i_{13,1}^{2}+ i_{23,2}^{2}\right), \label{Second}
\end{equation}

Further investigations are needed to establish a strict theoretical relationship between the microscopic structure of considered reinforcements and the macroscopic parameters here introduced: it is indeed well known that the second gradient parameters are intrinsically related to a characteristic length $L_{c}$ which is, in turn, associated to the micro-structural properties of considered materials. Many identification methods have been introduced to relate the macroscopic second gradient parameter to the microscopic properties of the considered medium, e.g. see  \cite{alibert2003truss,seppecher2011exotic}. Suitable multi-scale methods as the one introduced in \cite{nadler2006multiscale} may be generalized to be applied to the present case. Moreover, the description of the considered system at the microscopic scale may exploit some of the results proposed in \cite{atai1997nonlinear,haseganu1996equilibrium,steigmann1992equilibrium}.

\section{Numerical model and results\label{sec:Numerical-Model}}

In this section, the implemented FEM and the consequent results for the simulation of deep-drawing are shown.

The structure of this section will be as follows:
\begin{itemize}
\item in the first subsection, the numerical implementation of the contact interaction between the testing machine and the interlock is presented,
\item in the second subsection the proposed constrained shape functions, that will be here called \textit{augmented continuity shape functions} and used to perform a numerical simulation of the presented second gradient model, will be defined,
\item in the third subsection, the influence of the second gradient parameter $\alpha$ on the onset of wrinkles during the simulation of the deep-drawing is studied,
\item the fourth subsection presents some observed the mesh-dependency result for the first gradient model, when using linear shape functions,
\item the final subsection shows the first and second gradient solutions as functions of the adopted mesh, when using the \textit{augmented continuity shape functions}. It is concluded that second gradient simulations are not significantly affected by the choice of the mesh, provided that the size of the elements is sufficiently small.
\end{itemize}

\subsection{Modeling geometry and contact interaction between the mold and the reinforcement}

\begin{figure}[H]
	\begin{centering}	
		\includegraphics[width=10cm]{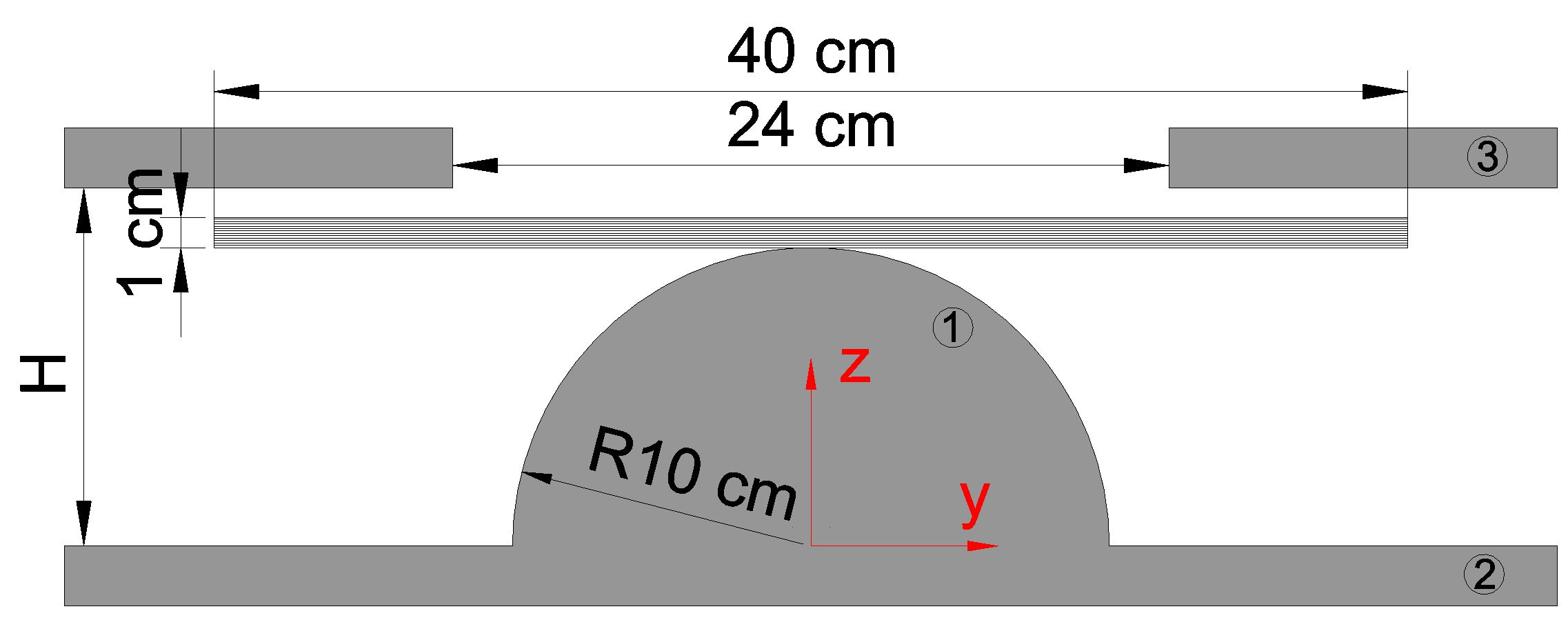} 
		\par\end{centering}
	\caption{Geometry of the model for a deep-drawing preforming with a hemispherical punch.\label{deepdrawinggeometry}}	
\end{figure}

The object of the paper is the simulation of the deep-drawing process performed on 2.5D composite interlocks. In particular, the focus will be on a hemispherical punch and dye system, as the one shown in Fig.\;\ref{deepdrawinggeometry}. In such test, a square dry woven composite interlock is formed by an hemispheric punch that, with the presence of a horizontal plane makes the deformed shape assume a double-curvature shape.

To implement the contact between the woven composite and the testing machine, a penalty function was introduced such that, to each interpenetration, it associates a stress $\mathbf{t}$ normal to the surface to be applied on the fabric, in formulas:
\begin{align}
\mathbf{t}=K_\mathrm{contact}\, \Delta\, \mathbf{n}, \label{Contact}
\end{align}
where $K_\mathrm{contact}$ is an opportune stiffness that is set to increase in each non-linear iteration so as to obtain the minimum interpenetration possible, $\Delta$ is the interpenetration between the woven fabrics and the machine, and $\mathbf{n}$ is the normal to the surface of the punch or of the die. The perpendicularity of the assumed stress is equivalent to assume that no friction is present between the experimental setup and the specimen. This could be considered a strong hypothesis, but the results seem to be qualitatively correct and this suffices for the purposes of this paper.

In the presented model, it was chosen to model the punch and the die as rigid bodies since they are supposed to have a stiffness of various orders of magnitude higher than the specimen. Thus, the position of the die and of the punch is known a priori in each of the steps of the test, leading to a much easier determination of the contact stresses in Eq.\,\eqref{Contact}. Indeed, considering as origin of the reference system the center of the basis of the hemispherical punch, the $\mathbf{Z}$ axis as the vertical loading direction and the current position $(x,y,z)$ of a point of the fabric, the interpenetrations between the woven fabric and the hemisphere, the lower plane and the die, respectively, can be expressed as:
\begin{align}
\Delta_{1}=\mathrm{max}\big(R-\sqrt{x^2+y^2+z^2},0\big),\qquad \Delta_{2}=\mathrm{max}\big(-z,0\big),\qquad \Delta_{3}=\mathrm{max}\big(z-H+w_{0},0\big),
\end{align}
where $R$ is the radius of the hemispherical punch, $H_0$ and $w_0$ are the initial position and the  applied displacement of the die in the considered step, respectively. The direction of the resulting stress is radial for the hemispherical punch and vertical for the plane and the die. Therefore, considering once again the center of the hemispherical punch as the origin of the reference system, it is possible to write:
\begin{align}
\mathbf{n}_{1}=\frac{(x,y,z)}{\sqrt{x^2+y^2+z^2}},\qquad\qquad \mathbf{n}_{2}=(0,0,1),\qquad\qquad \mathbf{n}_{3}=(0,0,-1),
\end{align}
The resulting stresses $\mathbf{t}_{1}$ and $\mathbf{t}_{2}$ were applied to the lower surface of the specimen, while $\mathbf{t}_{3}$ was applied to the upper surface. Finally, the resulting contact stresses can be written as:
\begin{align}
\mathbf{t}_{1}&=K_\mathrm{contact}\,\mathrm{max}\left(\frac{R}{\sqrt{z^2+y^2+z^2}}-1,0\right)\,(x,y,z),\nonumber\\ 
\mathbf{t}_{2}&=K_\mathrm{contact}\,\mathrm{max}\big(-z,0\big)\,(0,0,1),\\[0.8em]
 \mathbf{t}_{3}&=K_\mathrm{contact}\,\mathrm{max}\big(z-H+w_{0},0\big)\,(0,0,-1).\nonumber
\end{align}

\subsection{Augmented continuity shape functions}
As it is well known, the usual steps followed in finite element models are as follows:
\begin{enumerate}
	\item Discretize the considered body $\Omega$ with a mesh. Each of the obtained sub-bodies $\Omega_e$ is called an element.
	\item In each element $\Omega_e$, introduce a suitable number $N_n$ of local nodes $\{x^e_1,x^e_2,..,x^e_{N_n}\}$ which are needed to define a basis of polynomials $\{f^e_1,f^e_2,..,f^e_{N_n}\}$, of order n, which will be subsequently used to build-up the solution. Of course, the number of points $N_n$ to be introduced depends on the order of the chosen polynomials (e.g. for 1D elements $N_n=n+1$). The polynomials $f^e_i$ are given and are built in such a way that:
	\begin{align}
	\begin{cases}
	f^e_i(x^e_j)=0, \qquad j\neq i,\\
	f^e_i(x^e_i)=1, \qquad i,j\in {1,2,...,N_n},\\
	f^e_i=0, \qquad \text{outside the element }\Omega_e,
	\end{cases}
	\end{align}
	\item Look for a global solution that takes the form:
	\begin{align}
	u=\sum_{e=1}^{N_e} \sum_{i=1}^{N_n} a_i^e f_i^e,
	\end{align}
	where $N_e$ is the total number of elements $\Omega_e$ of the chosen discretization of $\Omega$, and $a_i^e$ are constants to be determined. Clearly, the number of constants $a_i^e$ depends both on the number of elements and on the order $n$ of the chosen polynomials $f_i^e$.
\end{enumerate}
The more common finite elements codes, such as the one used to perform the simulations in this paper, are built in such a way that the unknown constants are determined according to the following procedure:
\begin{enumerate}
	\item calculate some $a_i^e$ by means of the imposed boundary conditions (e.g. assigned displacement at a boundary node),
	\item calculate some others $a_i^e$ by imposing continuity of the field $u$ through the elements $\Omega_e$,
	\item calculate the remaining constants $a_i^e$ by means of a suitable minimization of the global action functional $\mathcal{A}$ associated to the energy W.
\end{enumerate}

Such elements are usually built to treat problems in classical elasticity where first gradient energies $W(\nabla u)$ are introduced. Note that the global continuity of displacement is imposed a priori in these elements.

The physical phenomena, object of this paper, typically show deformed shapes that exhibit continuity of displacement and also of strain (first derivatives of displacement). Indeed, from the experimental observations, it is possible to notice that, even if there are rapid changes of the strain within the specimen, they are always smoothened by the presence of transition layers allowing to pass from a value of strain to the other in a continuous way (see e.g. \cite{barbagallo2016bias,ferretti2014modeling,madeo2016continuum}).

To describe the considered phenomena, namely continuous displacements and rapid but continuous variation of strain, two ingredients are needed:
\begin{enumerate}
	\item  the introduction of a second gradient energy, e.g. of the type in Eq.\,\eqref{Second},
	\item  an adapted finite element which is able to assure the continuity of the first derivatives of displacement (class of continuity $C^1$).
\end{enumerate}

Such second point can be achieved at least in two ways:
\begin{enumerate}
	\item introducing a supplementary kinematical field $P:\Omega\rightarrow\mathbb{R}^3$, subsequently constrained to be related to first derivatives of displacement as $P\rightarrow(i_{12},i_{13},i_{23})$ e.g. using Lagrange multipliers or penalty methods (e.g. see \cite{madeo2016continuum,madeo2015thick}),
	\item keeping the same kinematics (only the standard displacement field) and try to force the finite element to grant continuity of first derivatives of displacement.
\end{enumerate} 

In this paper, the authors chose to implement this second way of granting continuity of strain by introducing third order Lagrangian polynomials which guarantee such \textit{augmented continuity} with a penalty energy at the element interfaces of the type\footnote{The term $[[\cdot]]$ denotes the jump of the quantity $\cdot\,$ at the interface between two elements $\Omega_e$.}:
\begin{align}
W_{\text{Interface}}=K_{\text{Penalty}}\,([[i_{12}]]^2+[[i_{13}]]^2+[[i_{23}]]^2).
\end{align}
This energy depends only on the discontinuity of the deformations $i_{12}$, $i_{13}$ and $i_{23}$ and it is, therefore, not sufficient to render the entire $\nabla u$ continuous. Nonetheless, the derivatives of the deformations $i_{12}$, $i_{13}$ and $i_{23}$ are the only ones appearing in the presented second gradient energy and, therefore, are the only ones on which the continuity is imposed. 

Whit this workaround it was possible to obtain almost continuous deformations and, thus, to implement directly a second gradient 3D model. The possibility of adding an energy on the interface between mesh elements is not always granted, but it is possible in COMSOL$^{\circledR}$, the software used for the simulations presented in the present paper.

\subsection{Influence of the second gradient on the wrinkling}

The model here presented implements the \textit{augmented continuity shape functions} in a COMSOL$^{\circledR}$ finite element model. The energy considered was the sum of the first gradient energy presented in Eq.\,\eqref{First} and of the second gradient one given in Eq.\,\eqref{Second}, for which the directions of the fibers $\mathbf{D}_{1}$ and $\mathbf{D}_{2}$ were chosen to be parallel to the edges of the specimen. The first gradient parameters are the ones shown in the Tab.\;\ref{table: Par}, while various values of the second gradient parameter $\alpha$ were considered. It must be noted that, in the case $\alpha=0$, the model reduces to a first gradient model with the energy in Eq.\,\eqref{First}.

\begin{figure}[H]
	\begin{centering}
		\begin{tabular}{|c|c|}
			\hline&\\ $1^{st}$ gradient model ($\alpha=0\ N$)& $\alpha=0.1\ N$ \\
			\includegraphics[width=7cm]{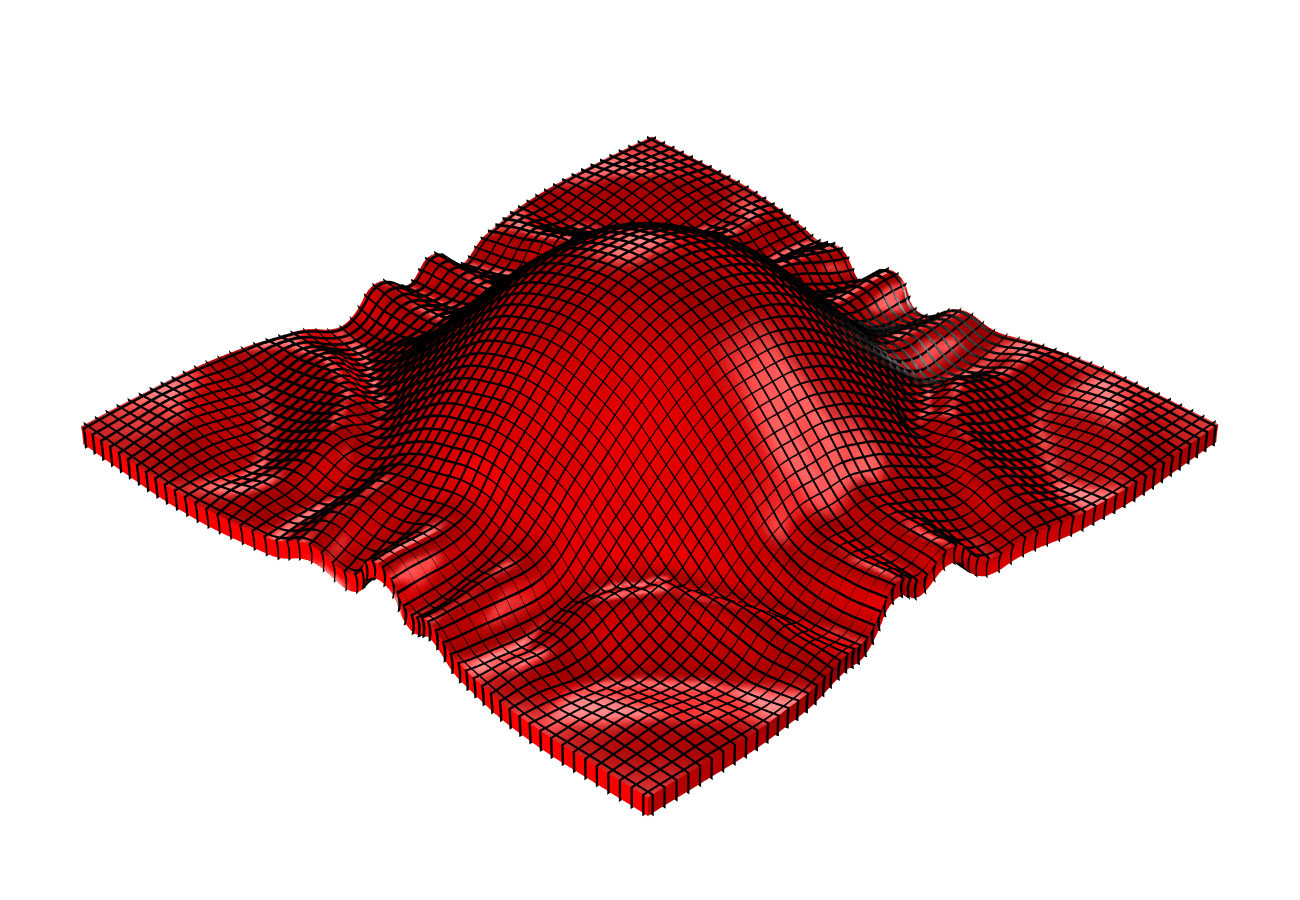}  & \includegraphics[width=7cm]{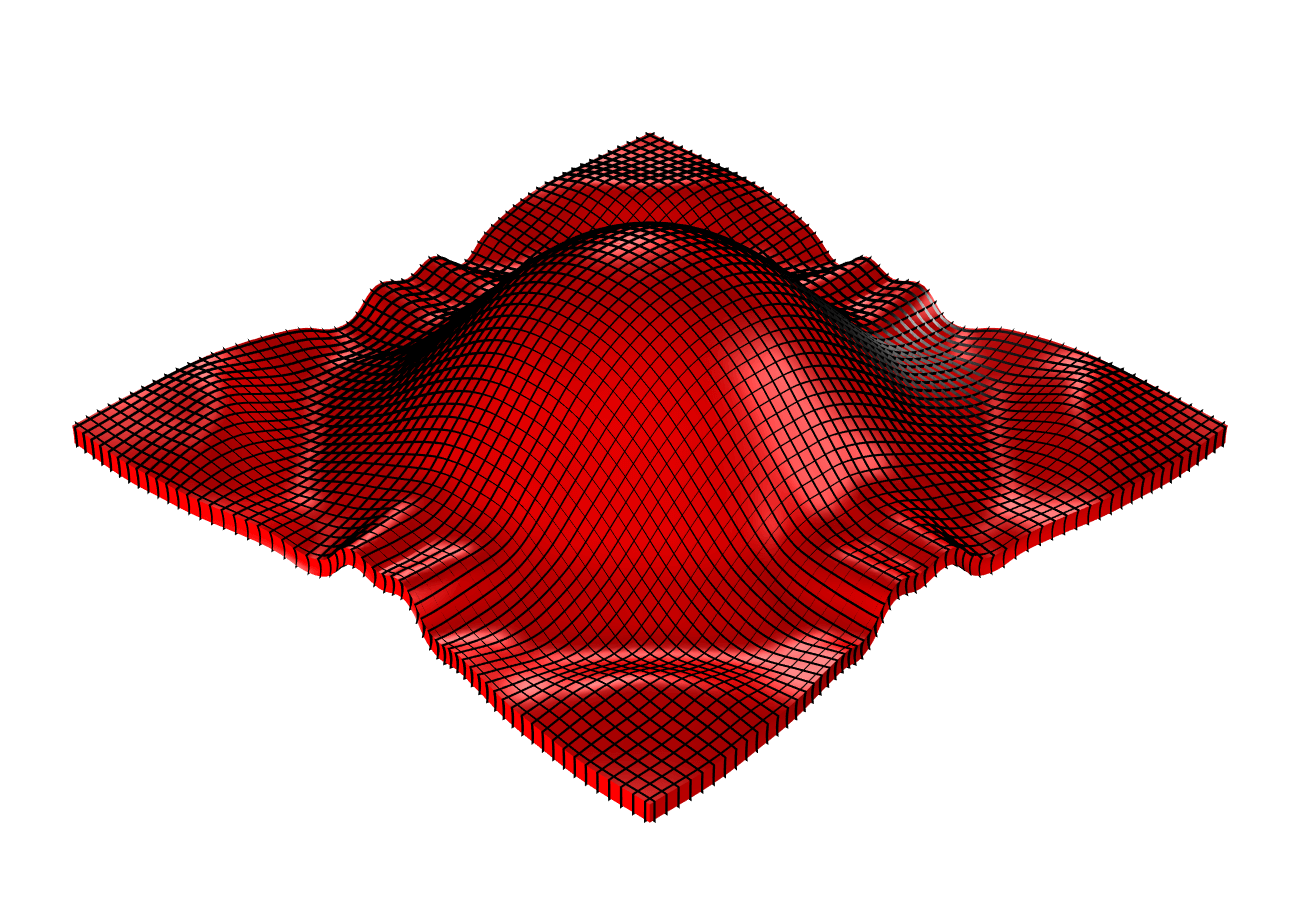}\\
			\hline&\\ $\alpha=1\ N$& $\alpha=10\ N$\\
			\includegraphics[width=7cm]{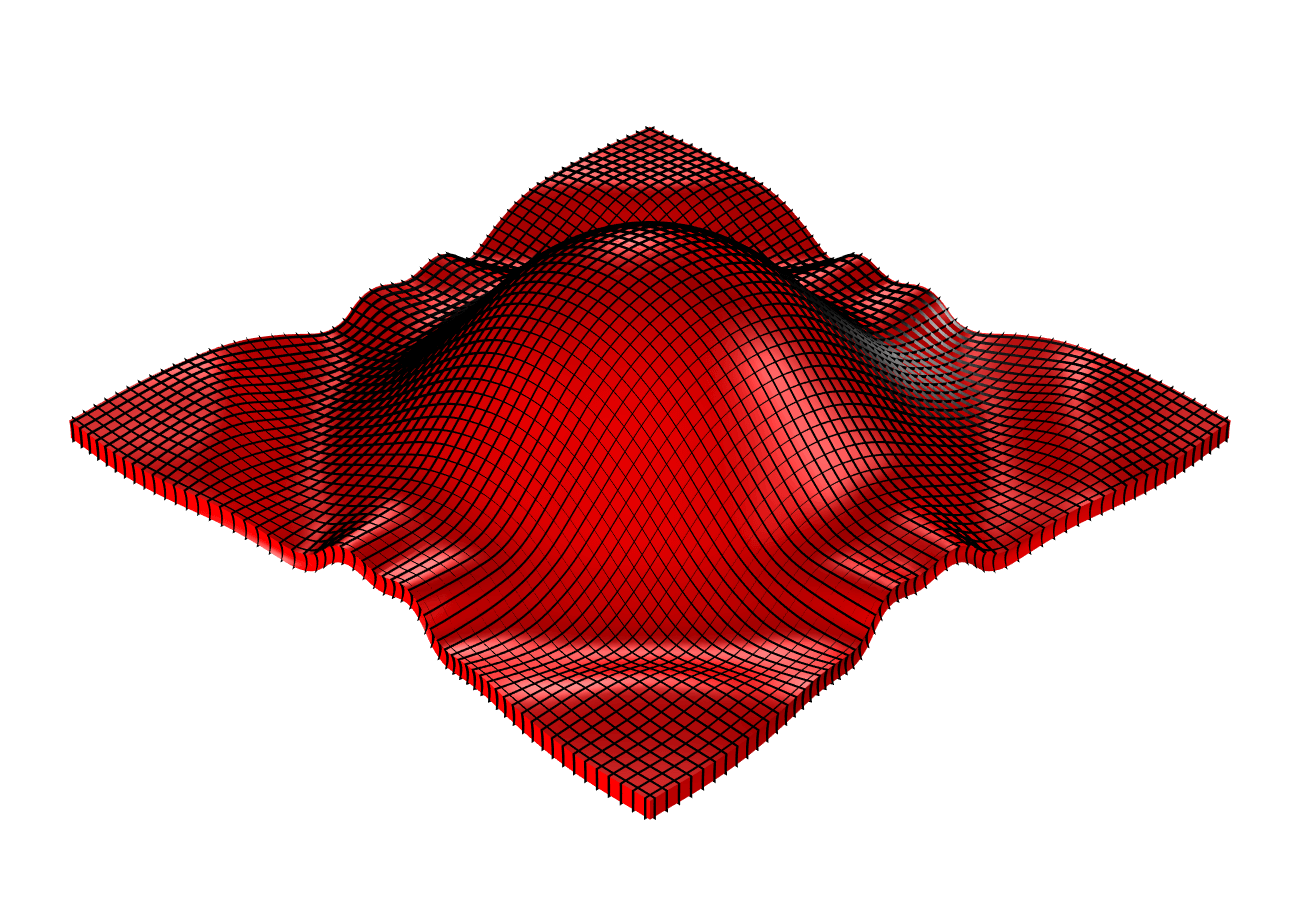}  & \includegraphics[width=7cm]{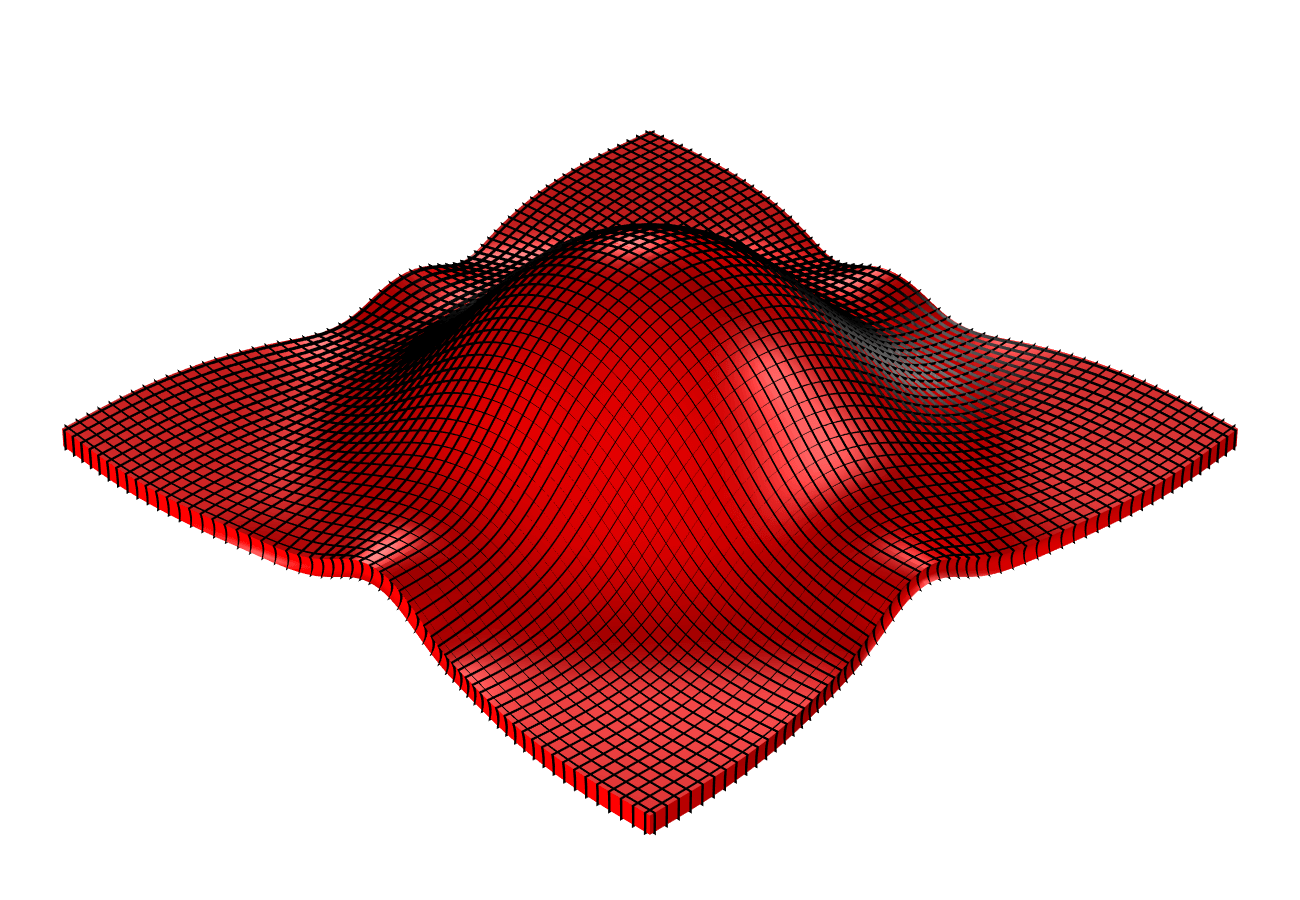}\\
			\hline
		\end{tabular}
		\par\end{centering}
	
	\caption{Dependence of the solution on the second gradient parameter $\alpha$.\label{SecGrad}	}	
\end{figure}

The results, obtained for $\alpha=0,0.1,1,10\,N$, are shown in Fig.\;\ref{SecGrad} for an imposed displacement of 90\% of the punch's radius (9 cm). In the first gradient case, it is possible to notice the presence of a significant number of wrinkles in the fibers direction causing a considerable out-of-plane curvature of the fibers. Instead, the insertion of a second gradient energy depending on the curvature generates a tendency to reduce the wrinkling effect. If the value of $\alpha$ increases to the value $10\,N$ all the secondary wrinkling disappears and the only principal wrinkle remaining is due to the natural evolution of the double curvature of the macroscopic configuration. 

As it will be shown in subsections \ref{sec:Mesh-1} and  \ref{sec:Mesh-2}, the first gradient model appears to be mesh-dependent even after the introduction of the \textit{augmented continuity shape functions} (see Fig. \ref{augm}) and it is, therefore, impossible to show a representative solution for this case (see subsections \ref{sec:Mesh-1} and  \ref{sec:Mesh-2}). Thus, it was chosen to show the deformed shape evaluated with the thinner used mesh, even if it is reasonable to assume that more wrinkles could appear for thinner meshes. On the other hand, even for small values of the second gradient parameter $\alpha$, a stabilization of the deformed configuration is obtained  (see Fig. \ref{augm})  and the deformed shape presented can be considered a representative solution, as it will be shown in subsections \ref{sec:Mesh-2}.

The possibility of controlling the onset and evolution of wrinkling during the deep-drawing simulation via the introduction of a constitutive parameter could be of great use in the prevision of the material behavior in view of structure design. It must be reminded that the presented simulations are relative to an experimental test that is meant for the characterization of the material constitutive properties. It is, therefore not enough to correctly describe the experimental results but the final goal is to predict the behavior of the woven fabric in generic engineering applications.

An issue that has to be covered is the determination of the second gradient parameter via experimental testing. Considering the proposed simple energy, it could be possible to heuristically choose $\alpha$ in order to have a qualitative description of the wrinkling phenomenon during a test such as the one proposed here. Furthermore, there are several experimental phenomena whose description would be useful for the calibration of a second gradient energy. During a Bias Extension Test, it is possible to observe the formation of some shear boundary layers the description of which can be used to calibrate the second energy parameters, as shown in \cite{ferretti2014modeling}. In the case of a Bias Extension Test on strongly unbalanced fabrics, the bending stiffness of the fibers can lead to some macroscopic effects like the asymmetric deformed shape analyzed in \cite{barbagallo2016bias,madeo2016continuum}. Finally, the calibration of the second gradient parameter could be attempted via a three point bending of an interlock, as in \cite{madeo2015thick}. Which combination of these tests is best suited for the determination of the second gradient parameters is still to be decided, but it is important to have multiple observable effects so that it is possible to validate the chosen parameters. 

The results obtained in this paper are a confirmation of the great potential of the use of a second gradient model for the description of the wrinkling phenomenon and, more generally, of the behavior of composite materials. In the authors' opinion, the results presented in this paper and in \cite{barbagallo2016bias,ferretti2014modeling,madeo2016continuum,madeo2015thick} are starting to clearly show how a second gradient model can be a potential solution for most of the issues relative to the description of the behavior of dry woven fibrous composite.

\subsection{Some considerations concerning mesh-dependency of the performed simulations}
\subsubsection{First gradient model with linear shape functions\label{sec:Mesh-1}}

As stated above, the results obtained via a first gradient model appear to be mesh-dependent, due to the non-stability of the wrinkling description. The aim of this subsection is to present this issue in the case of a classical first gradient implementation and, hence, to show the stabilization effect obtained with the insertion of a second gradient energy.

\begin{figure}[H]
	\begin{centering}
		\begin{tabular}{|c|c|c|c|}
			\hline&&&\\ \rotatebox{90}{\hspace{1cm}Mesh}&
			\includegraphics[trim=0 30 0 80,width=4.8cm]{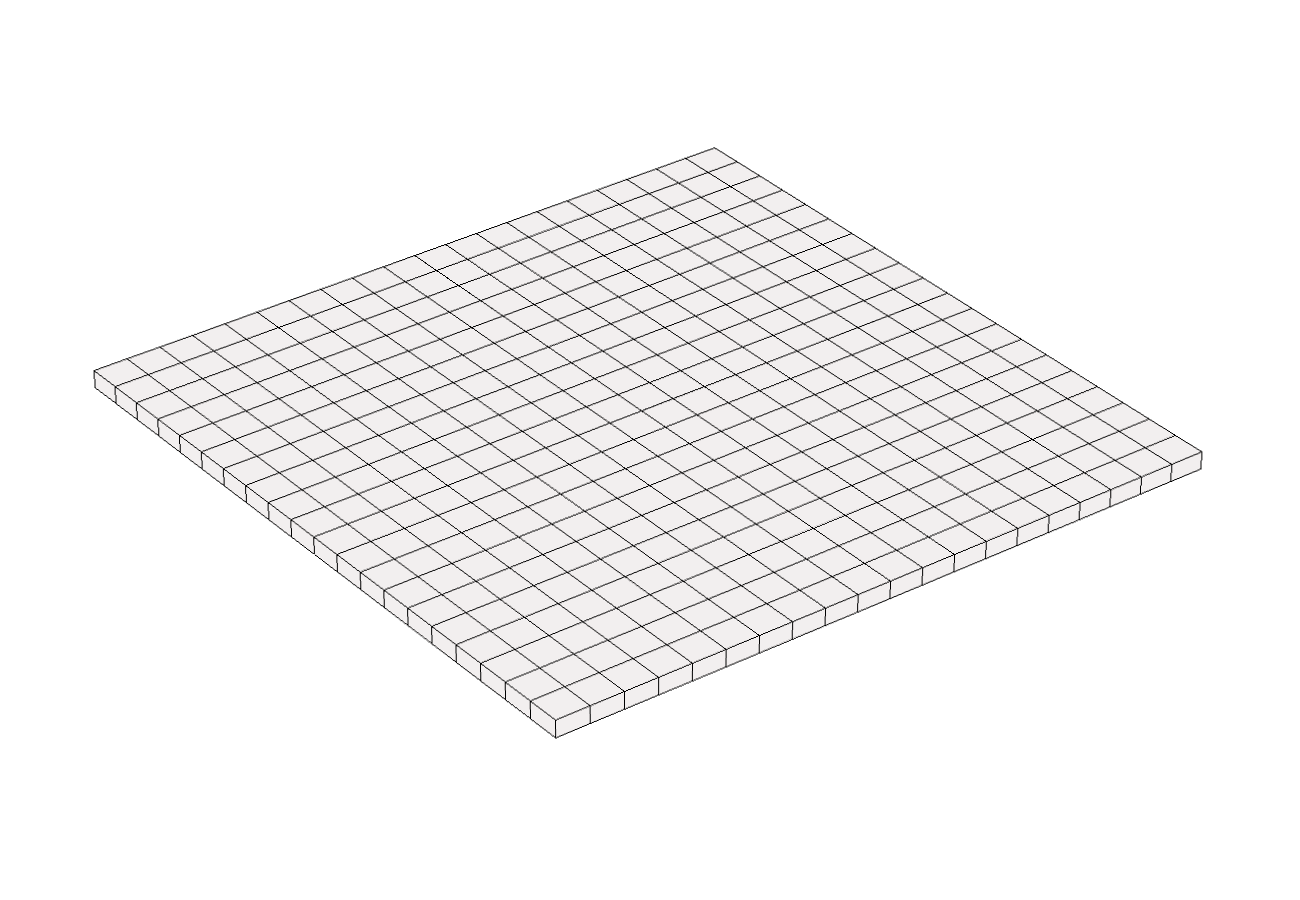}  & \includegraphics[trim=0 30 0 80,width=4.8cm]{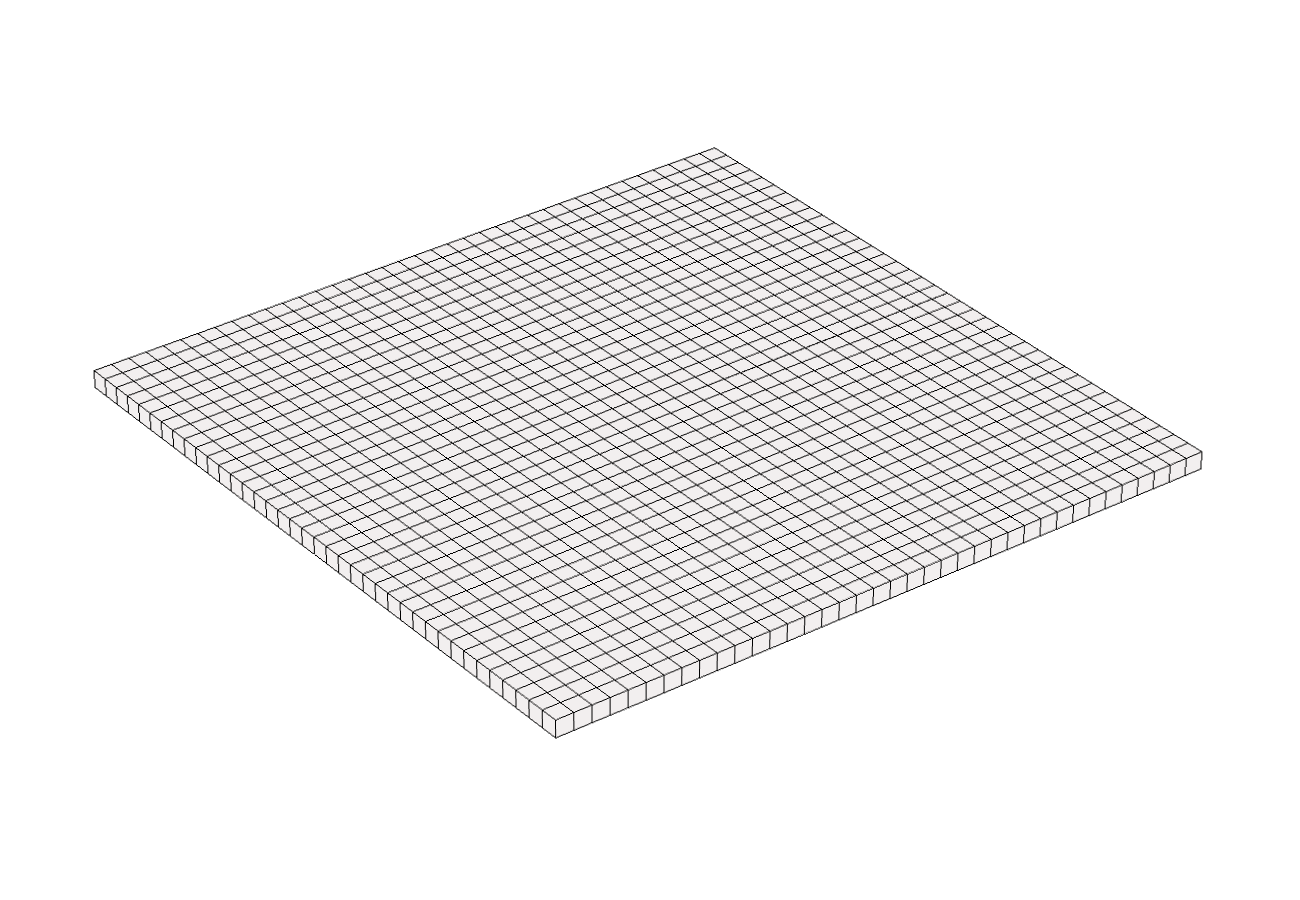} & \includegraphics[trim=0 30 0 80,width=4.8cm]{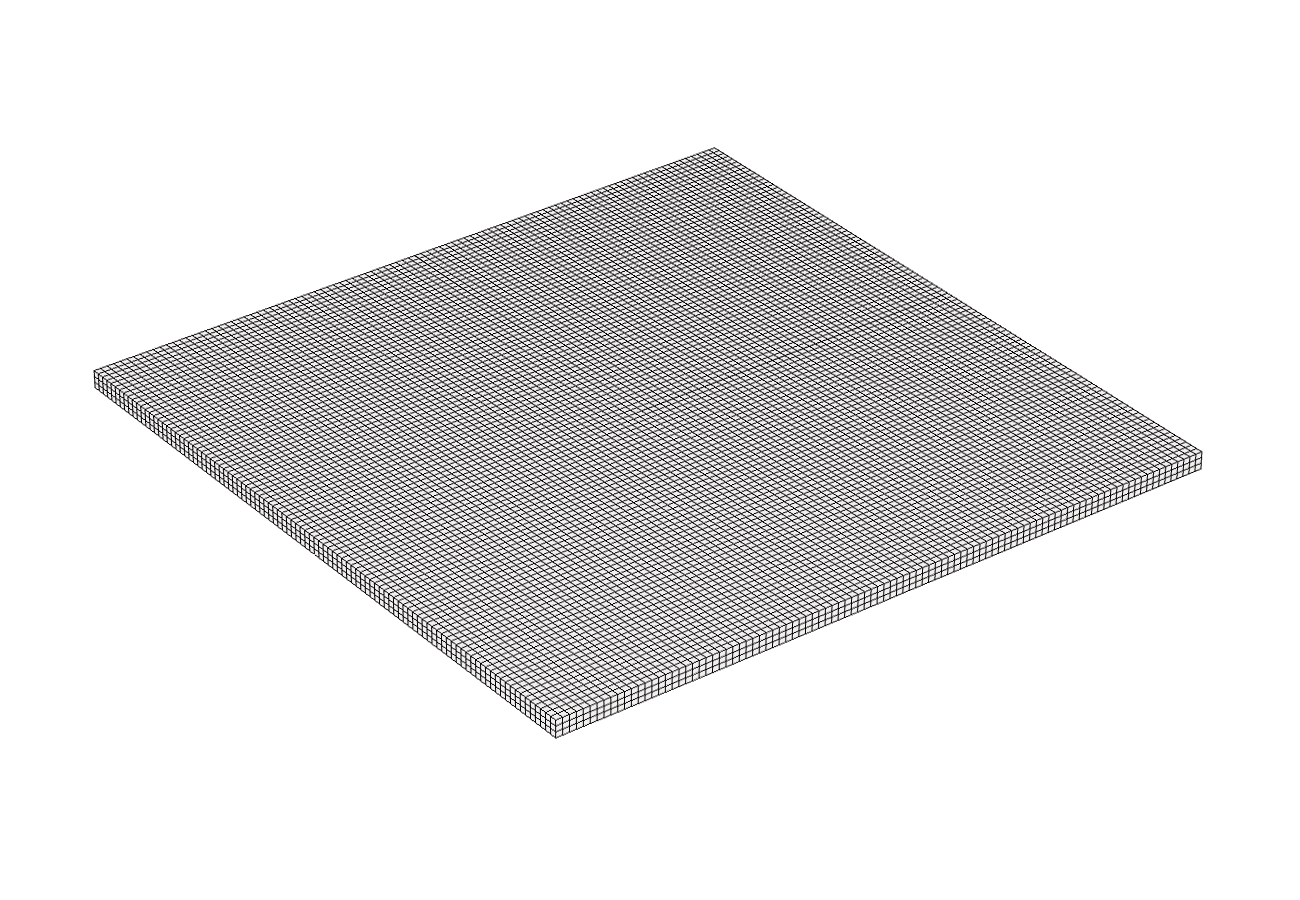} 
			\\\hline&&&\\\rotatebox{90}{\hspace{0.2cm}Deformed shape}&
			\includegraphics[trim=0 30 0 80,width=4.8cm]{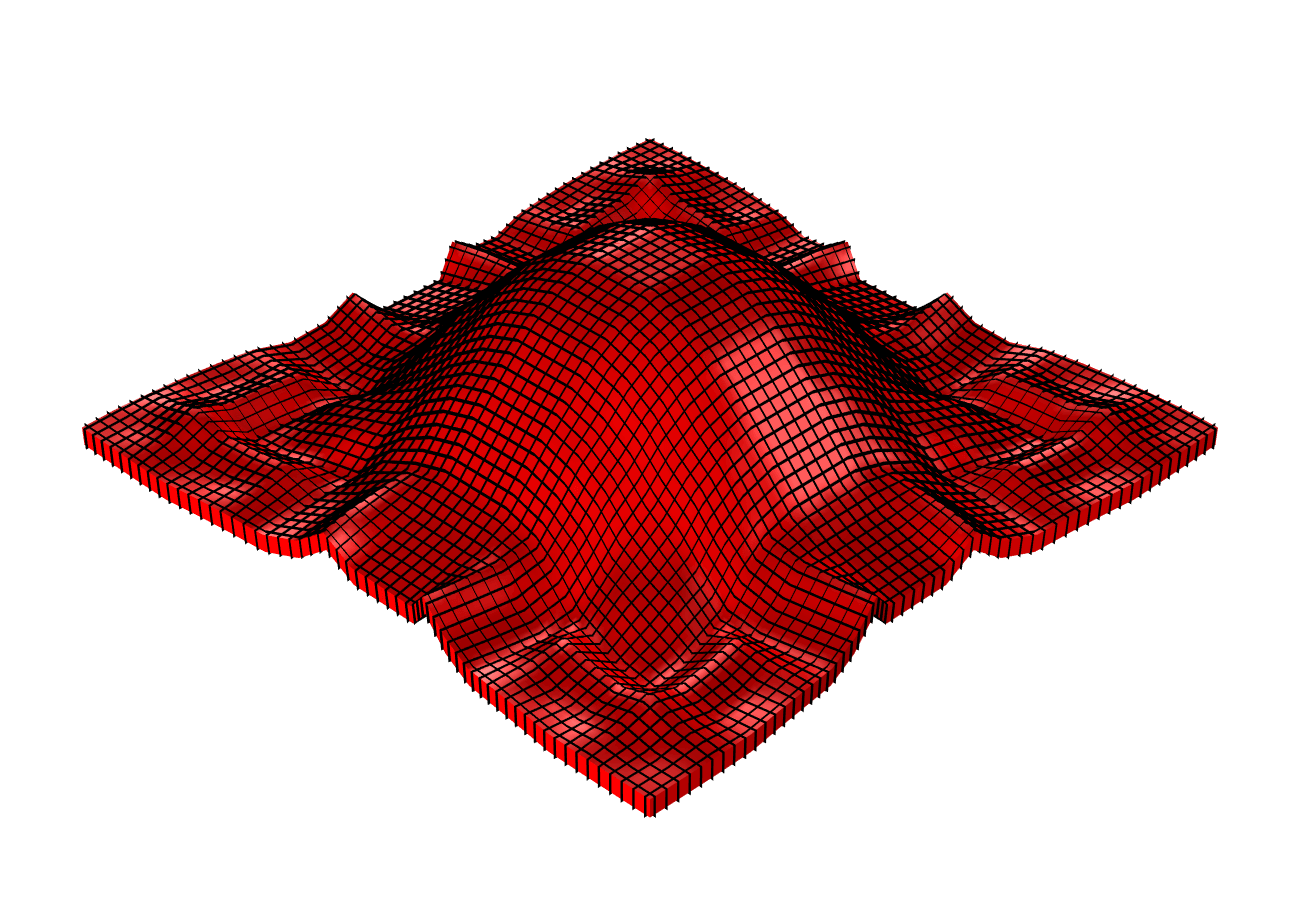}  & \includegraphics[trim=0 30 0 80,width=4.8cm]{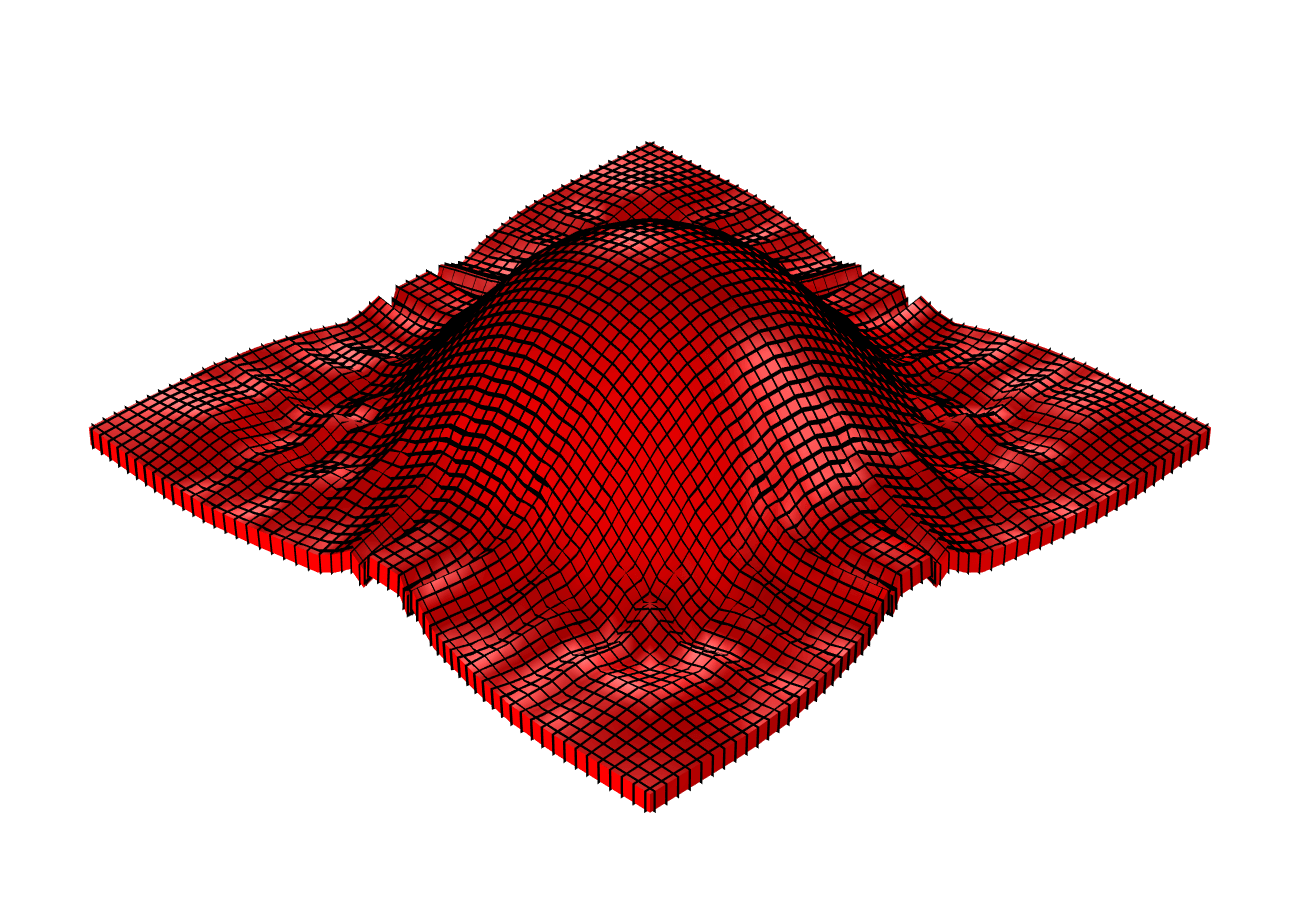} & \includegraphics[trim=0 30 0 80,width=4.8cm]{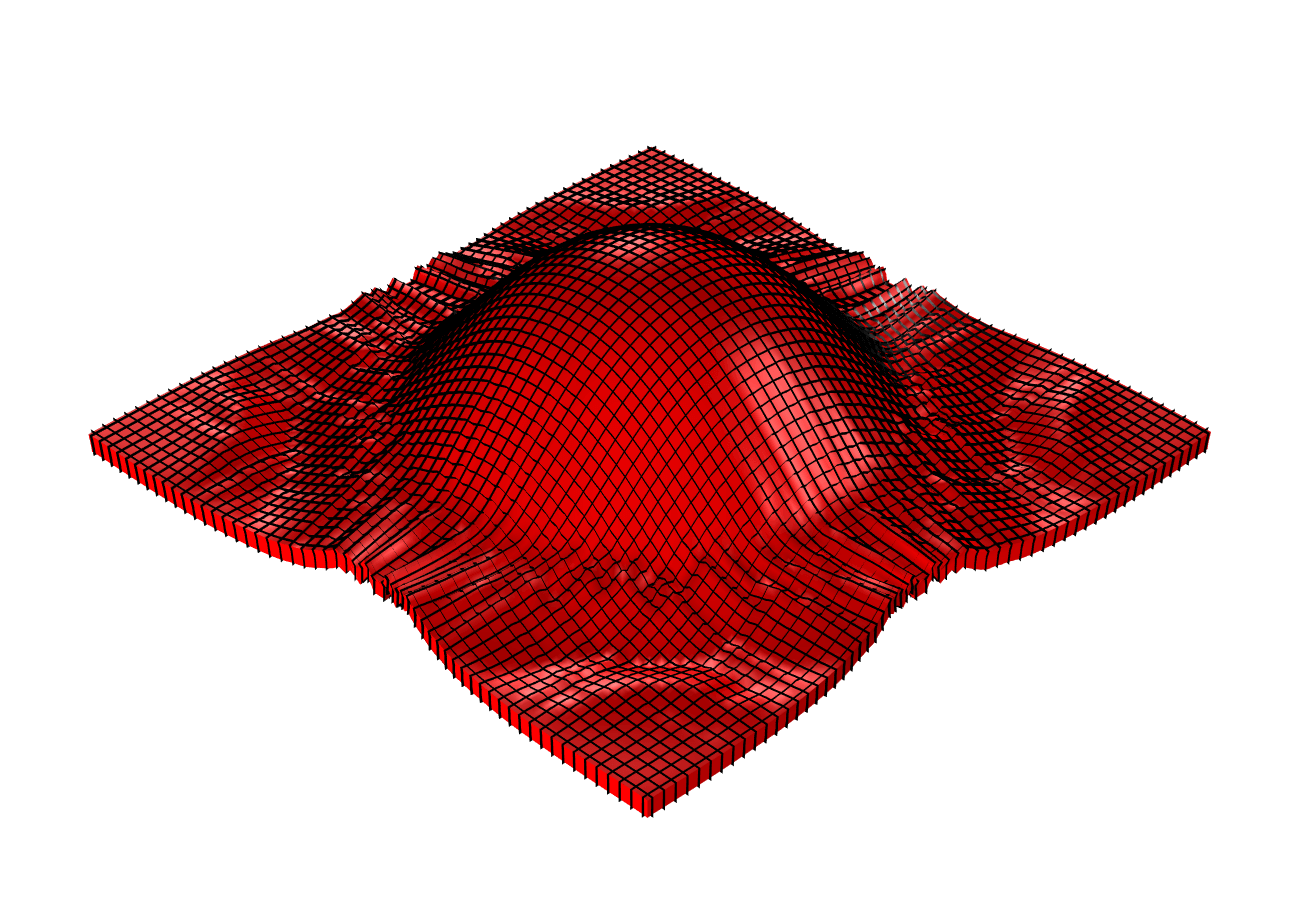} \\\hline
		\end{tabular}
		\par\end{centering}
	
	\caption{Solution of a first gradient model with linear shape functions and hexahedral meshes of different sizes.	\label{mapped}}	
\end{figure}

\begin{figure}[H]
	\begin{centering}
		\begin{tabular}{|c|c|c|c|}
			\hline&&&\\ \rotatebox{90}{\hspace{1cm}Mesh}&
			\includegraphics[trim=0 30 0 80,width=4.8cm]{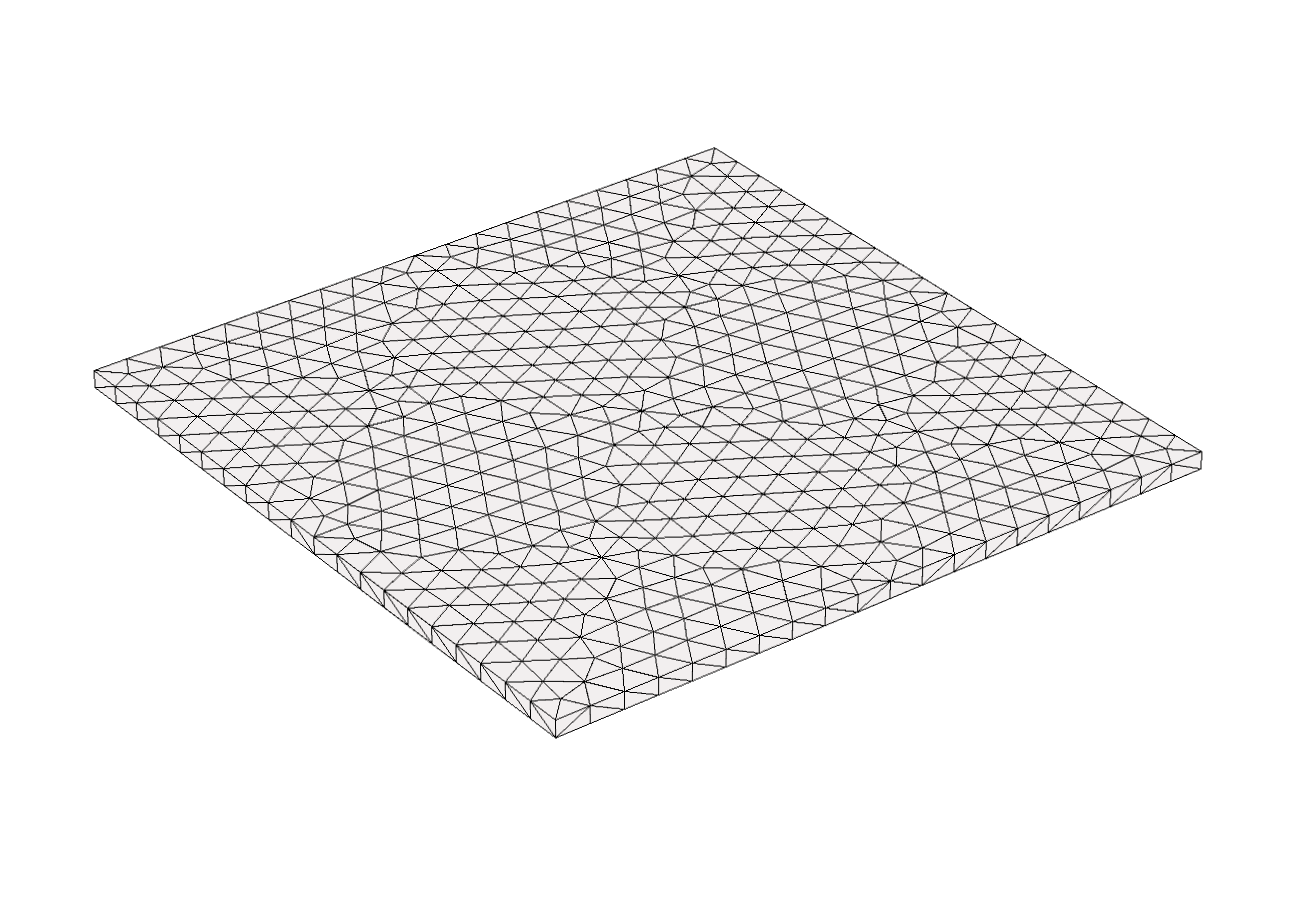}  & \includegraphics[trim=0 30 0 80,width=4.8cm]{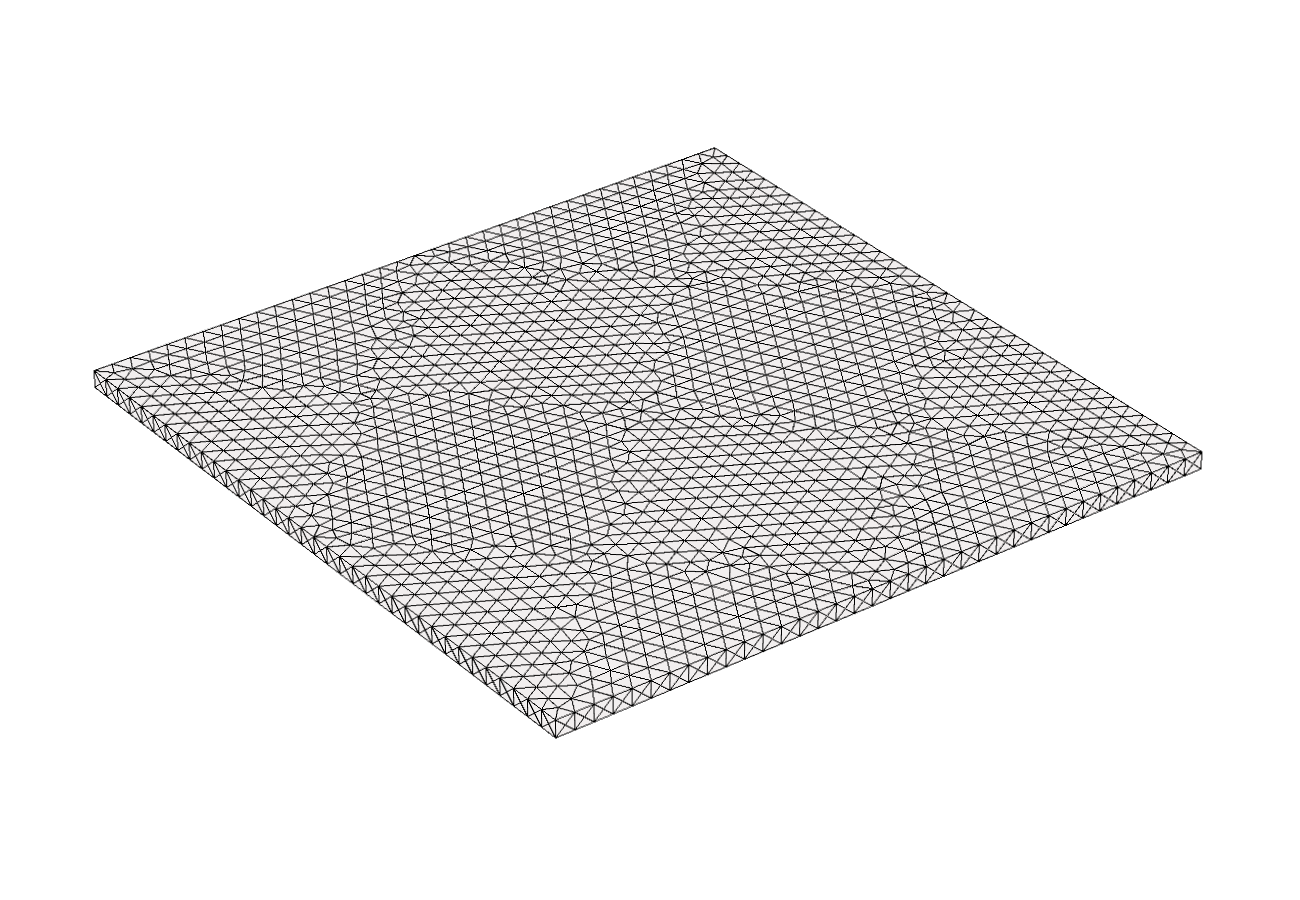} & \includegraphics[trim=0 30 0 80,width=4.8cm]{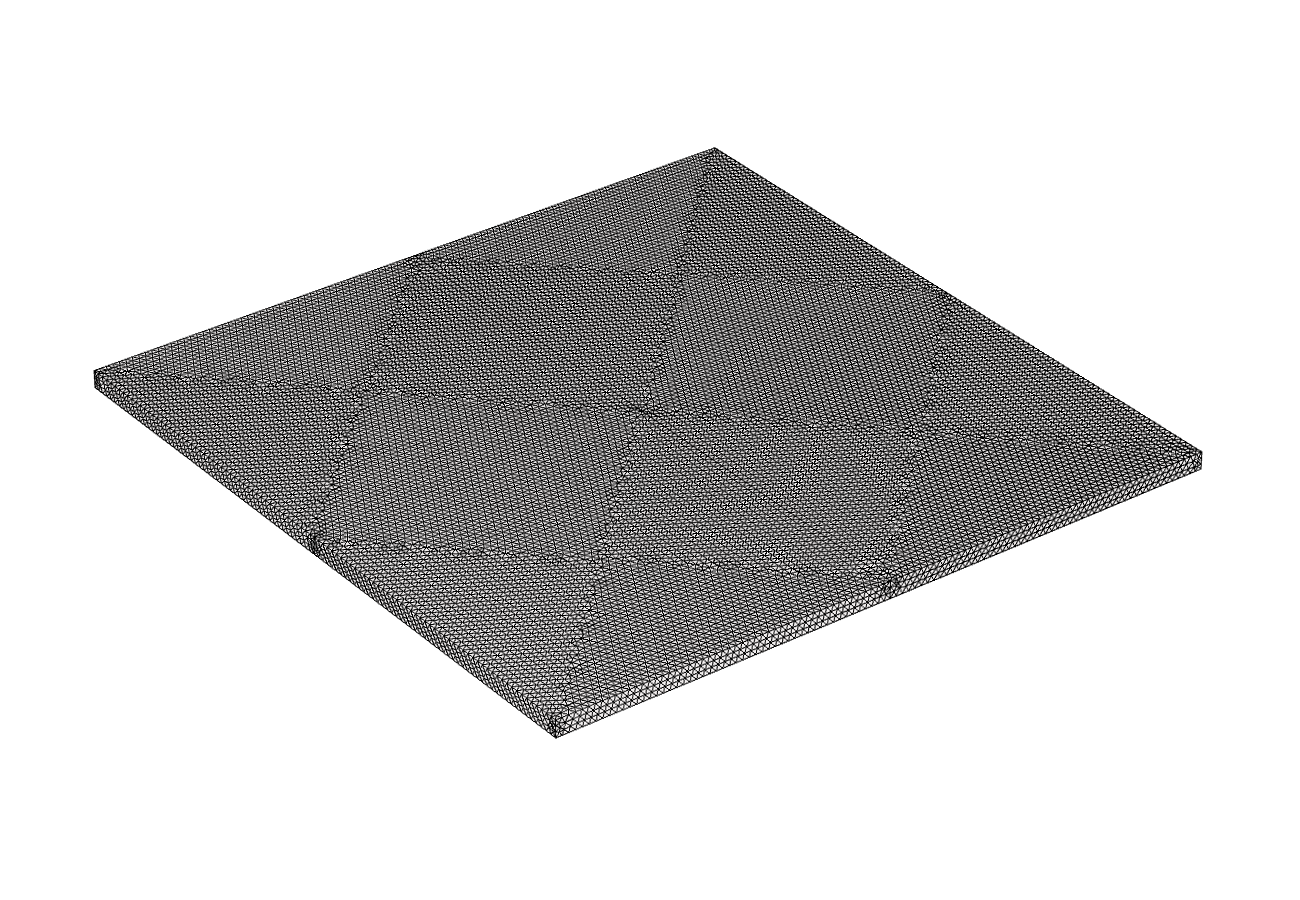} 
			\\\hline&&&\\\rotatebox{90}{\hspace{0.2cm}Deformed shape}&
			\includegraphics[trim=0 30 0 80,width=4.8cm]{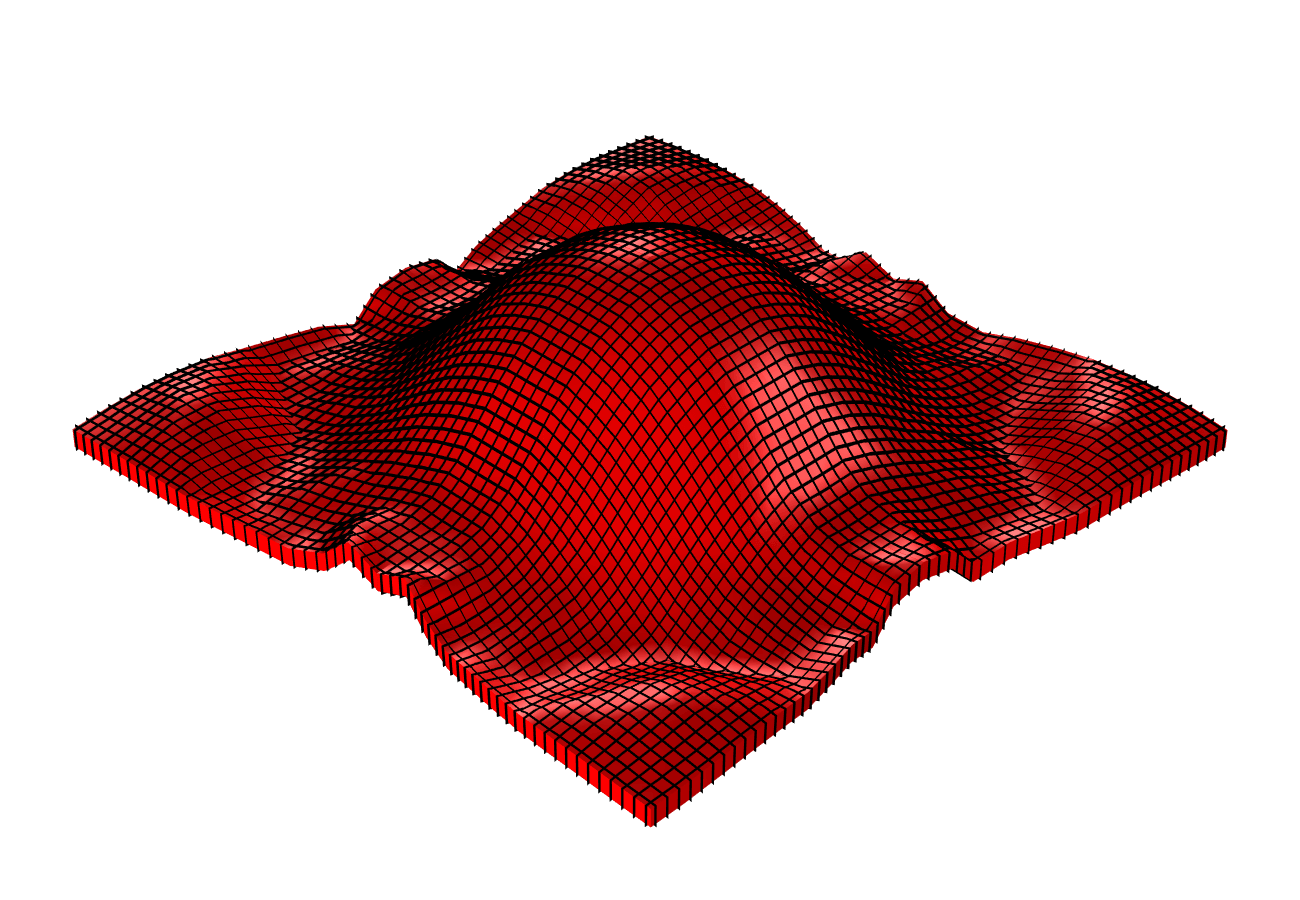}  & \includegraphics[trim=0 30 0 80,width=4.8cm]{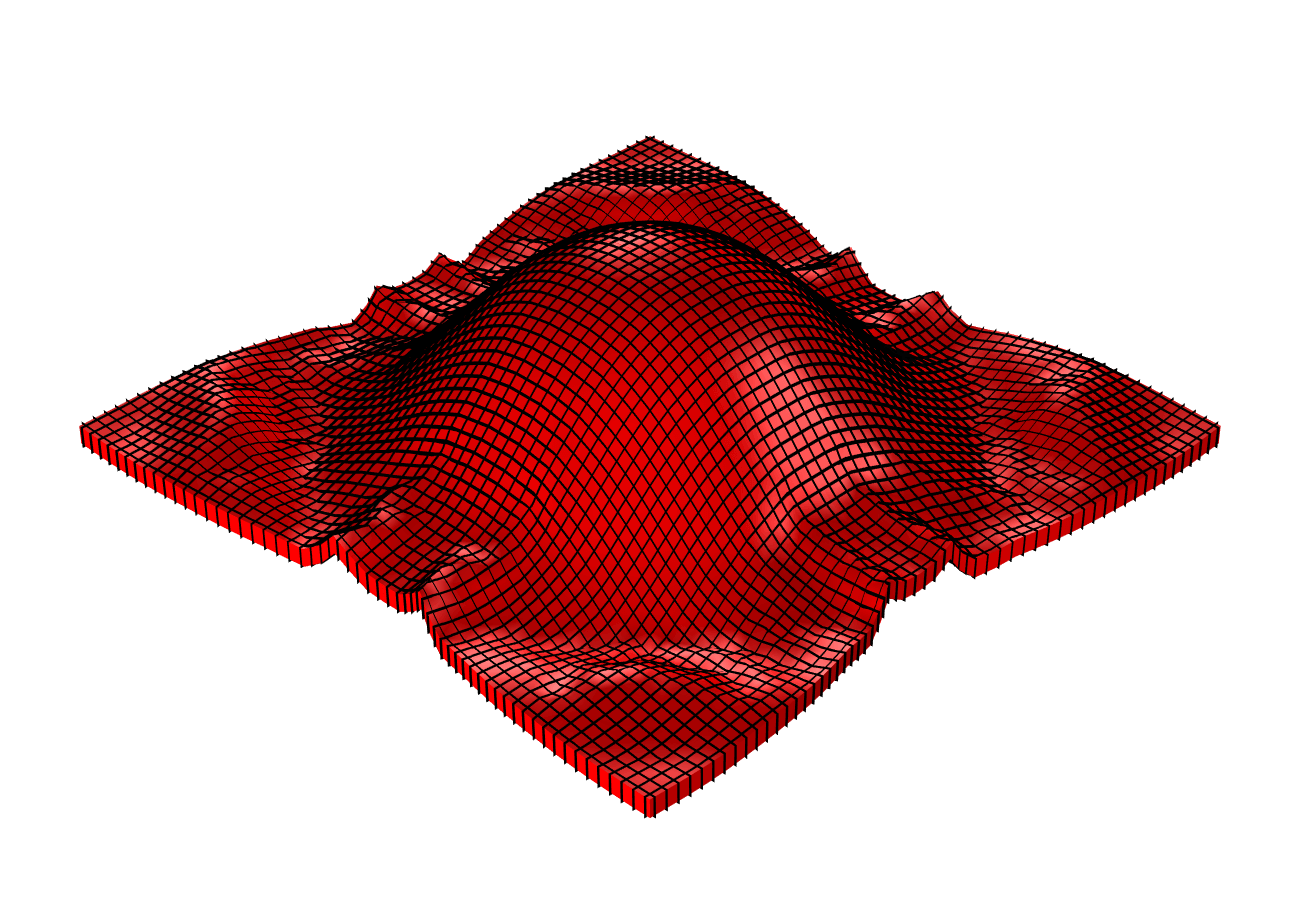} & \includegraphics[trim=0 30 0 80,width=4.8cm]{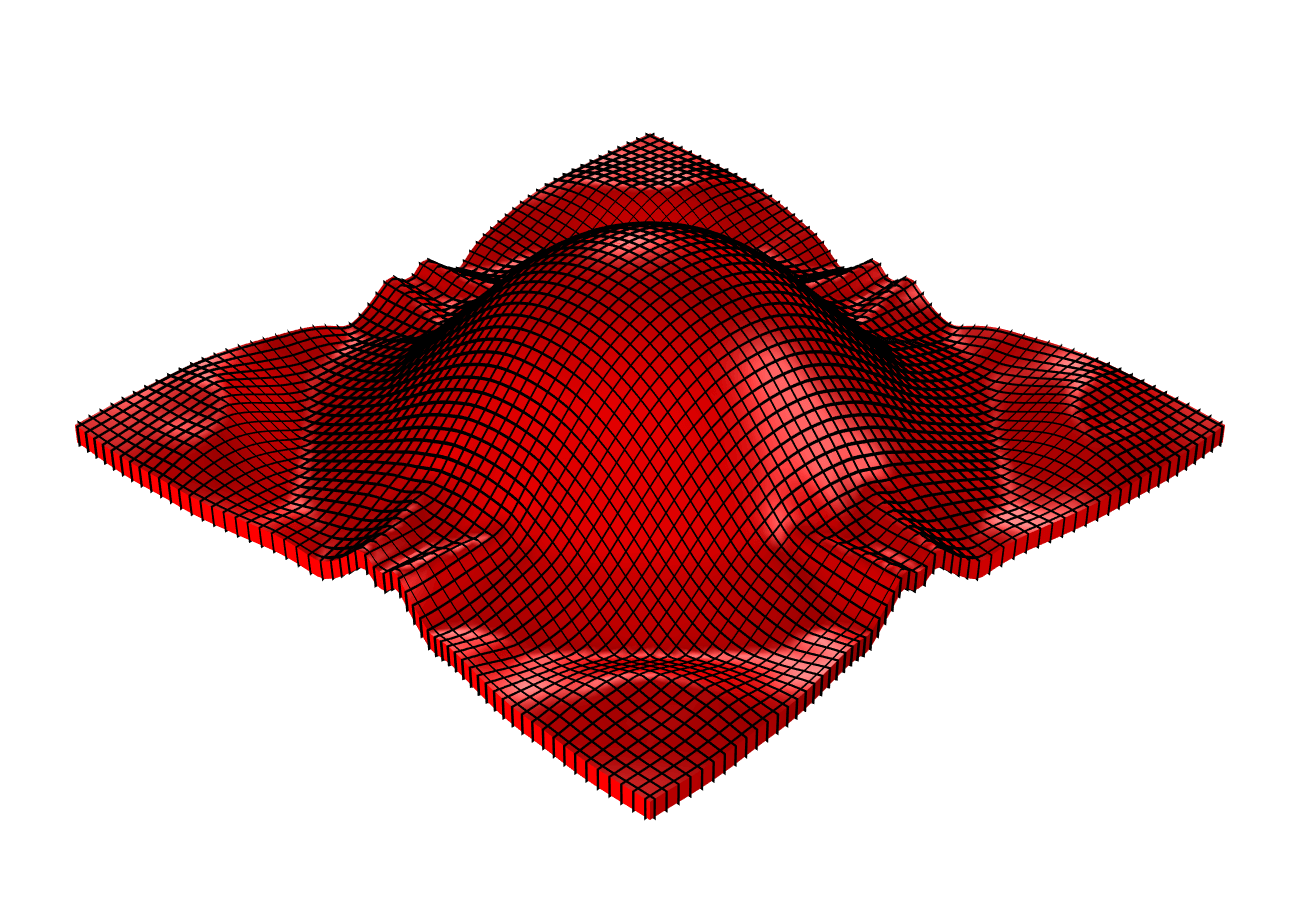} \\\hline
		\end{tabular}
		\par\end{centering}
	
	\caption{ Solutions for a first gradient model with linear shape functions and tetrahedral elements of different sizes.
		\label{tetra}	}	
\end{figure}

The \textit{augmented continuity shape functions} introduced for the second gradient model are momentarily discarded, so that it is possible to frame the mesh-dependency problem in a more traditional setting. In the simulations of this subsection, it was chosen to implement a model with the Lagrange linear shape functions. The study of the mesh-density's  influence on the first gradient solution is made with two types of mesh, namely:
\begin{itemize} 
	\item hexahedral meshes obtained sweeping quadrilateral meshes on the boundary over the thickness of the specimen (Fig.\;\ref{mapped}),
	\item tetrahedral elements (Fig.\;\ref{tetra}).
\end{itemize}
The same results cannot be obtained for the second gradient model because with such low continuity shape functions the insertion of a second gradient energy cannot be detected and, hence, it plays no actual role in the results. 

In Fig.\;\ref{mapped}, the hexahedral meshes and the resulting deformed shapes of the specimen are shown. It is important to remark that in this set of meshes the directions of the yarns $\mathbf{D}_{1}$ and $\mathbf{D}_{2}$, that are parallel to the edges of the specimen, coincide with the normals to the mesh interfaces. This property makes it possible to have a discontinuity on the derivatives in one of the fiber directions without losing the smoothness in the other direction. In other words, it is possible to form a wrinkle at the element interfaces for one set of fibers keeping the other set of fiber unaffected. This uncoupling can cause the formation of several wrinkles without interfering in other deformation mechanisms. Being the effect strongly related to the positioning and number of interfaces between meshes, it is not surprising that the result appears to be mesh-dependent. As a matter of fact, it is possible to see in Fig.\;\ref{mapped} how the increase in mesh-density seems to be connected to an increment in the number of wrinkles. 

Changing the type of mesh to tetrahedral elements as shown in Fig.\;\ref{tetra}, the improvement obtained in the stability is very clear. Despite the solution is once again mesh-dependent, the differences obtained in the output are much less with respect to the hexahedral mesh. The explanation for this result is that, in this case, the normals to the interfaces between the meshes do not always coincide with the direction of the fibers making the appearance of a wrinkling phenomenon at the interfaces more difficult. 

The conclusion on the results presented in this subsection is that a first gradient model with linear shape function may present unphysical wrinkling phenomena. It is still possible to obtain realistic results from such a model, but the dimension and the orientation of the elements should be carefully chosen to avoid unphysical wrinkling.

\subsubsection{First and second gradient models with augmented continuity shape functionss\label{sec:Mesh-2}}

\begin{figure}[H]
	\begin{centering}
		\begin{tabular}{|c|c|c|c|}
			\hline 		&&&\\ \rotatebox{90}{\hspace{1cm}Mesh}&
	\includegraphics[trim=0 60 0 60,width=4.8cm]{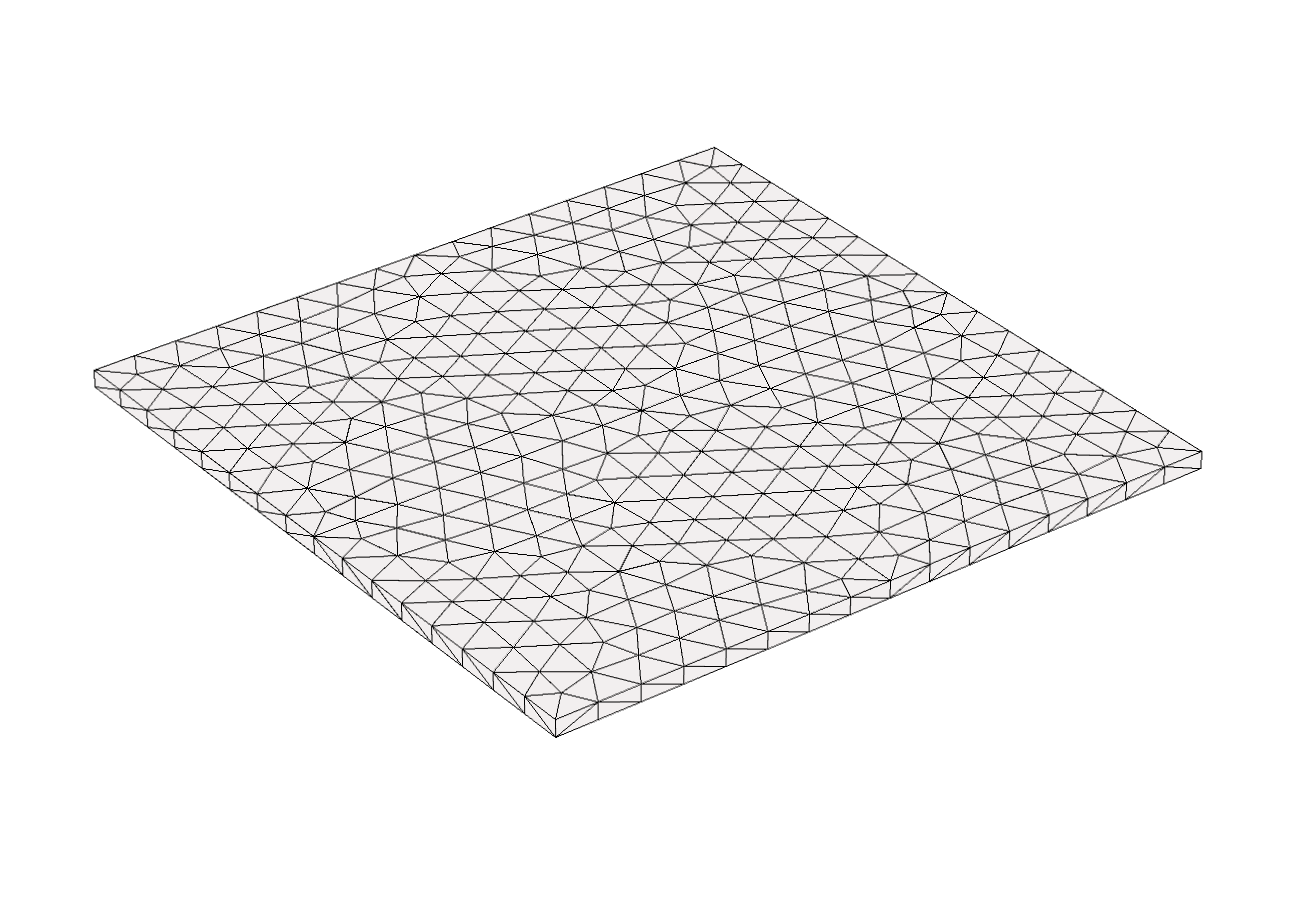}  & \includegraphics[trim=0 60 0 60,width=4.8cm]{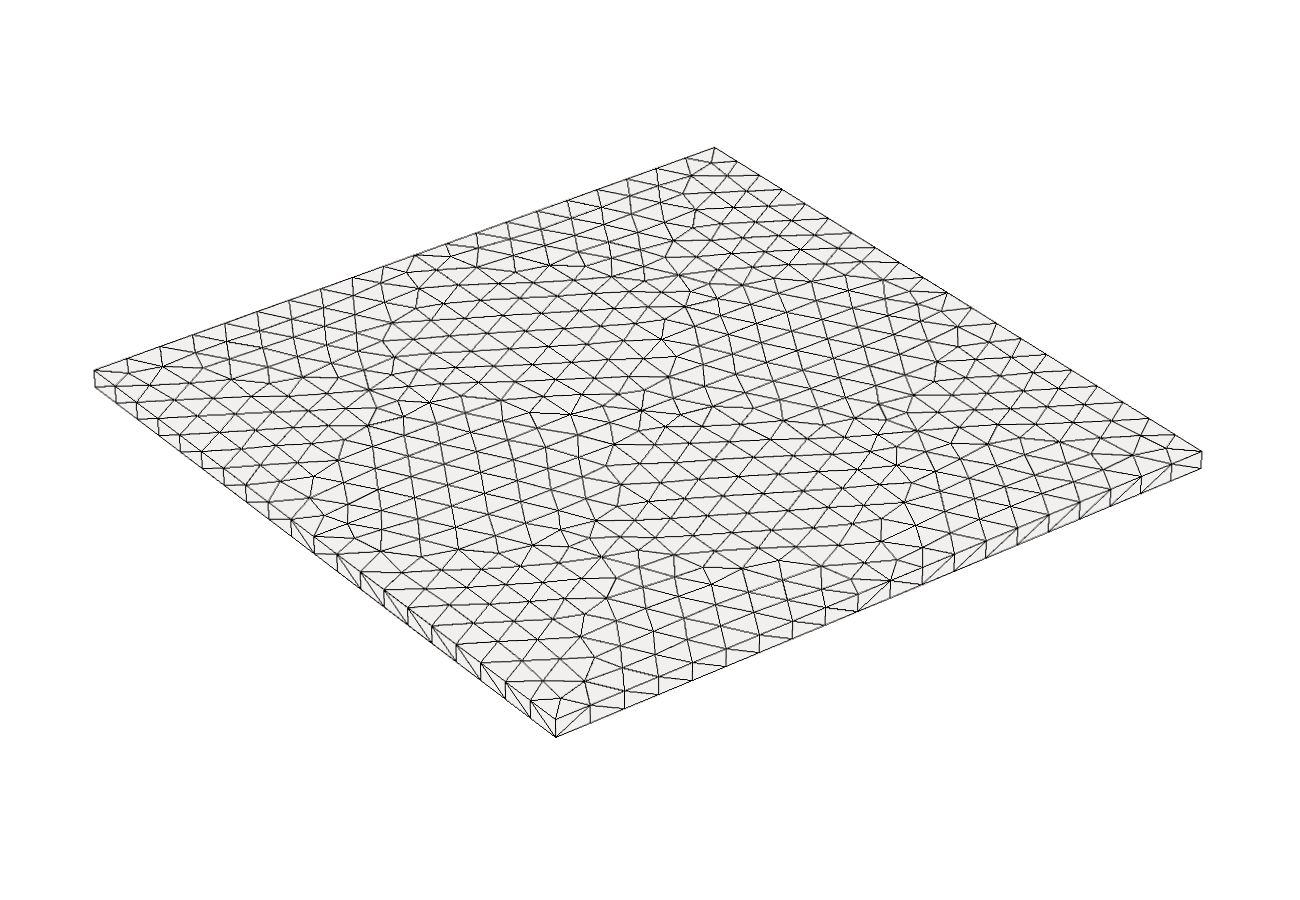} & \includegraphics[trim=0 60 0 60,width=4.8cm]{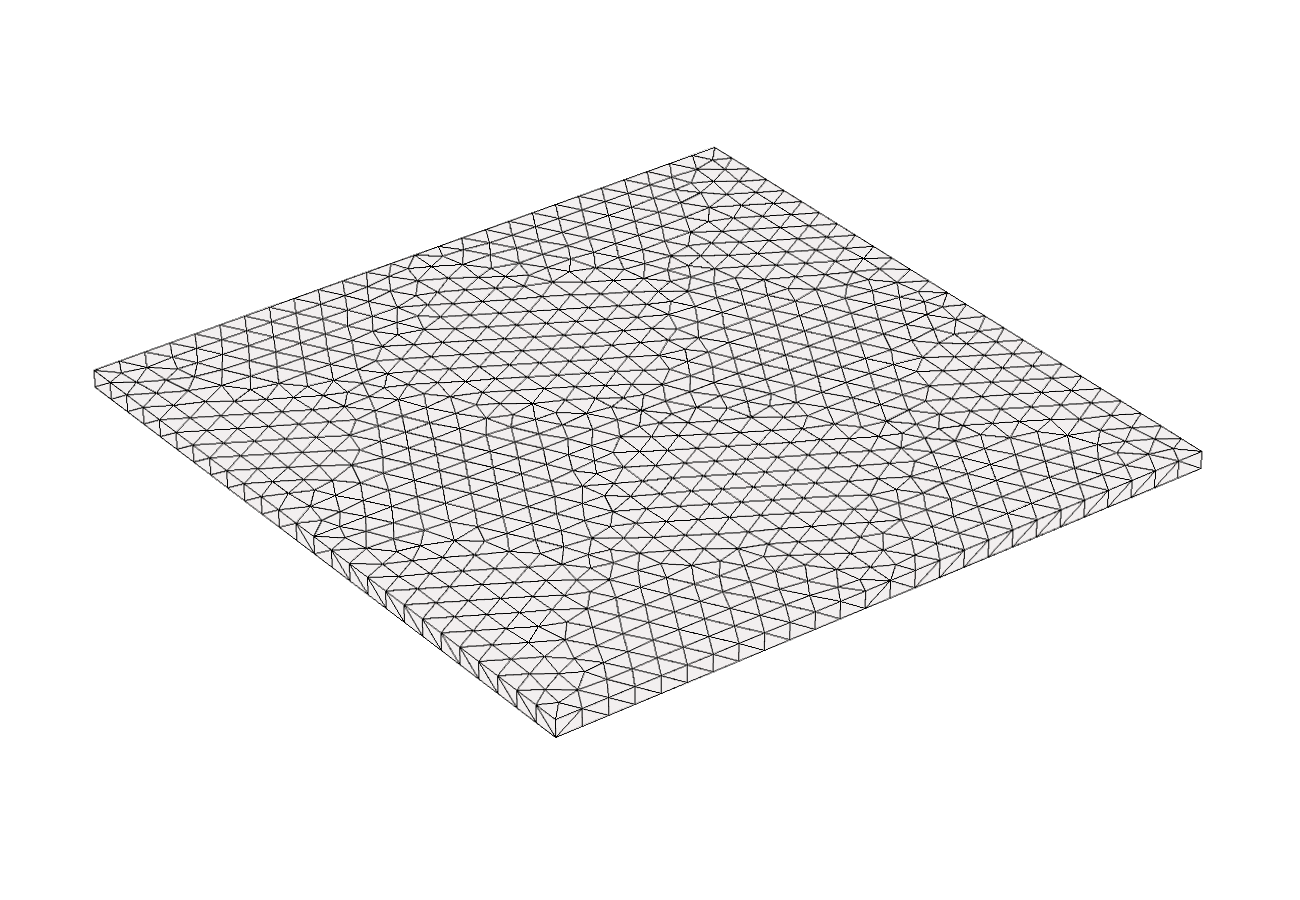} \\\hline&&&\\\rotatebox{90}{\hspace{0.5cm}$1^{st}$ gradient}&
			\includegraphics[trim=0 30 0 80,width=4.8cm]{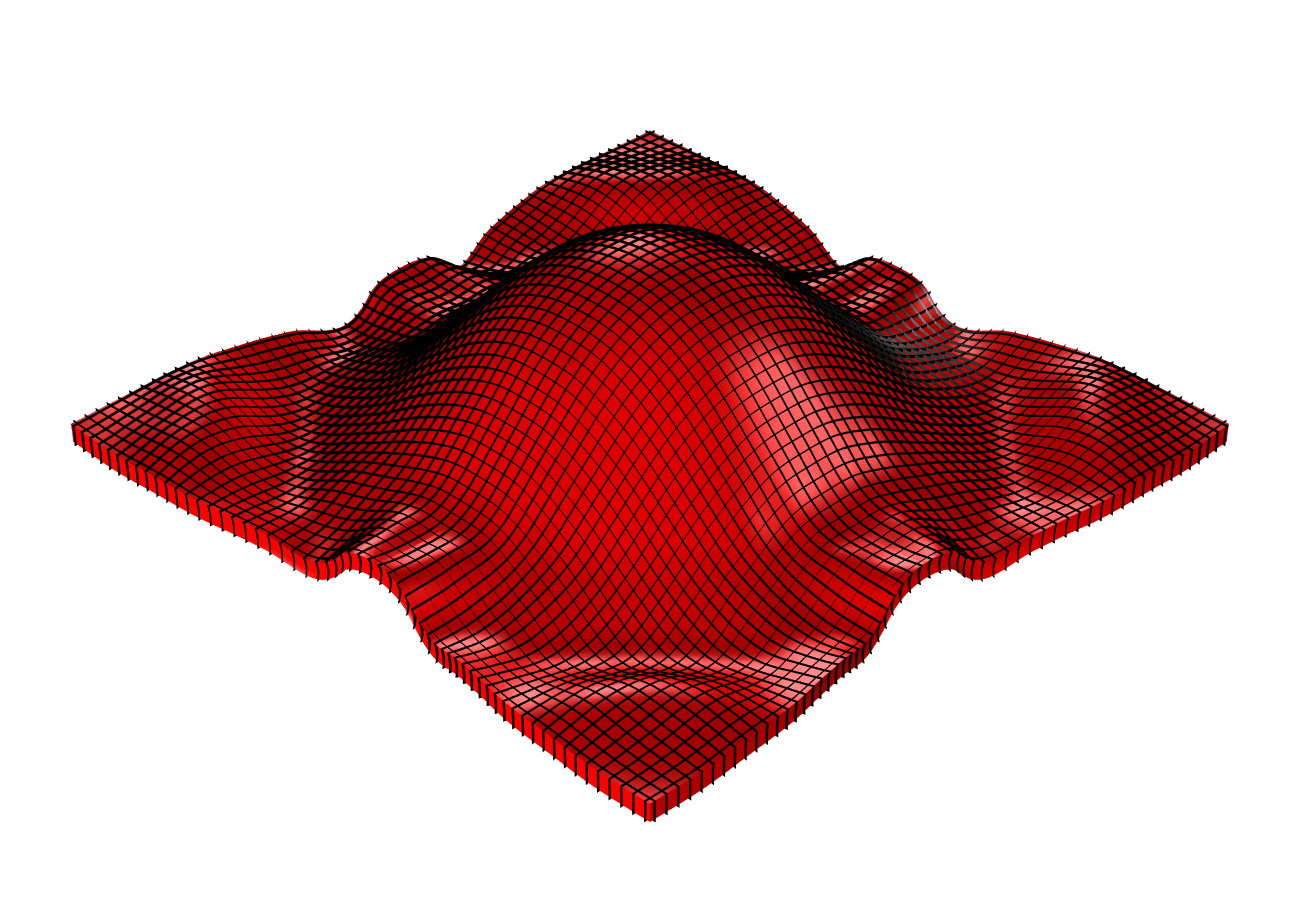}  & \includegraphics[trim=0 30 0 80,width=4.8cm]{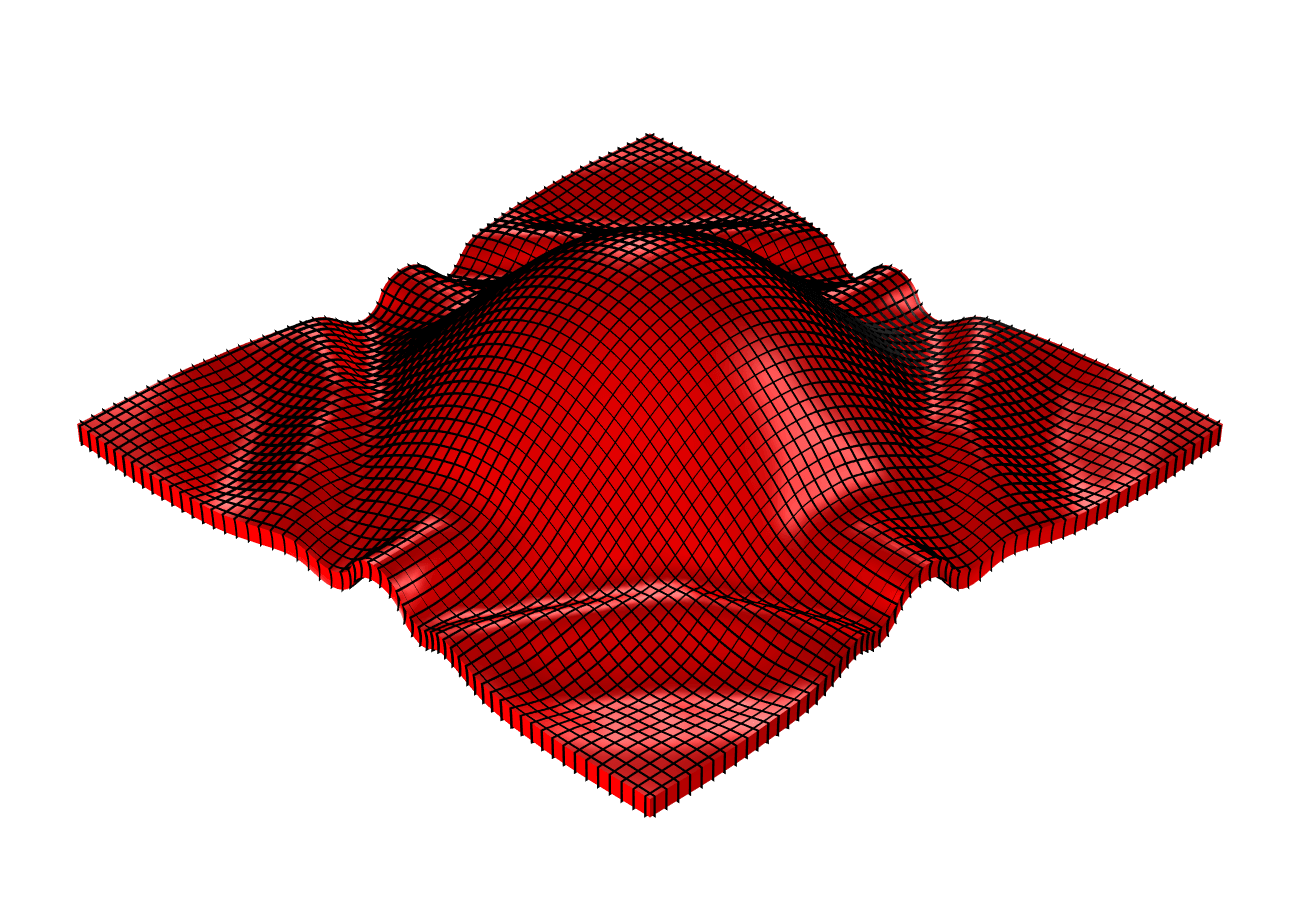} & \includegraphics[trim=0 30 0 80, width=4.8cm]{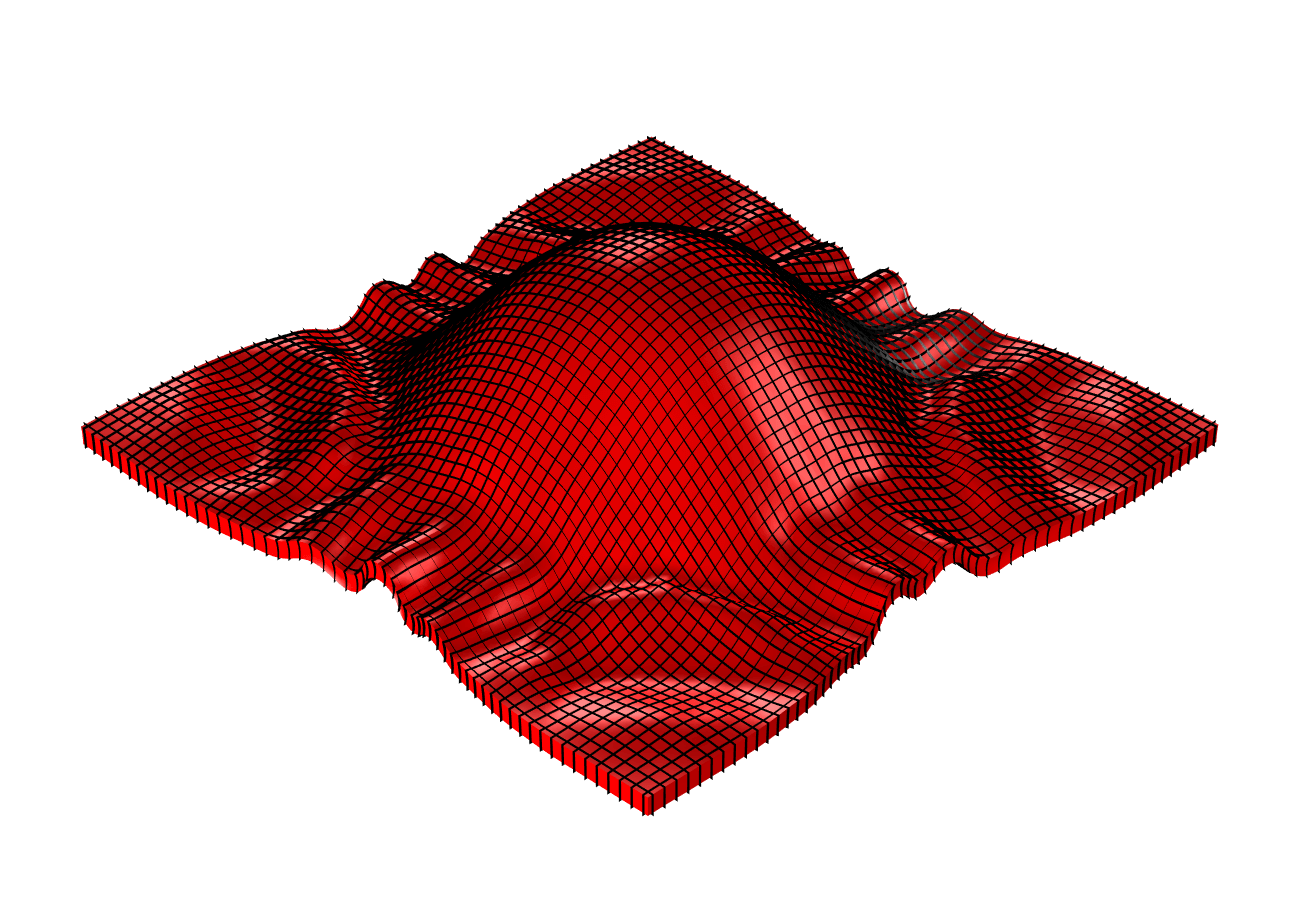} \\\hline&&&\\\rotatebox{90}{ \hspace{0.7cm}$\alpha=0.1\,N$}&
			\includegraphics[trim=0 30 0 80,width=4.8cm]{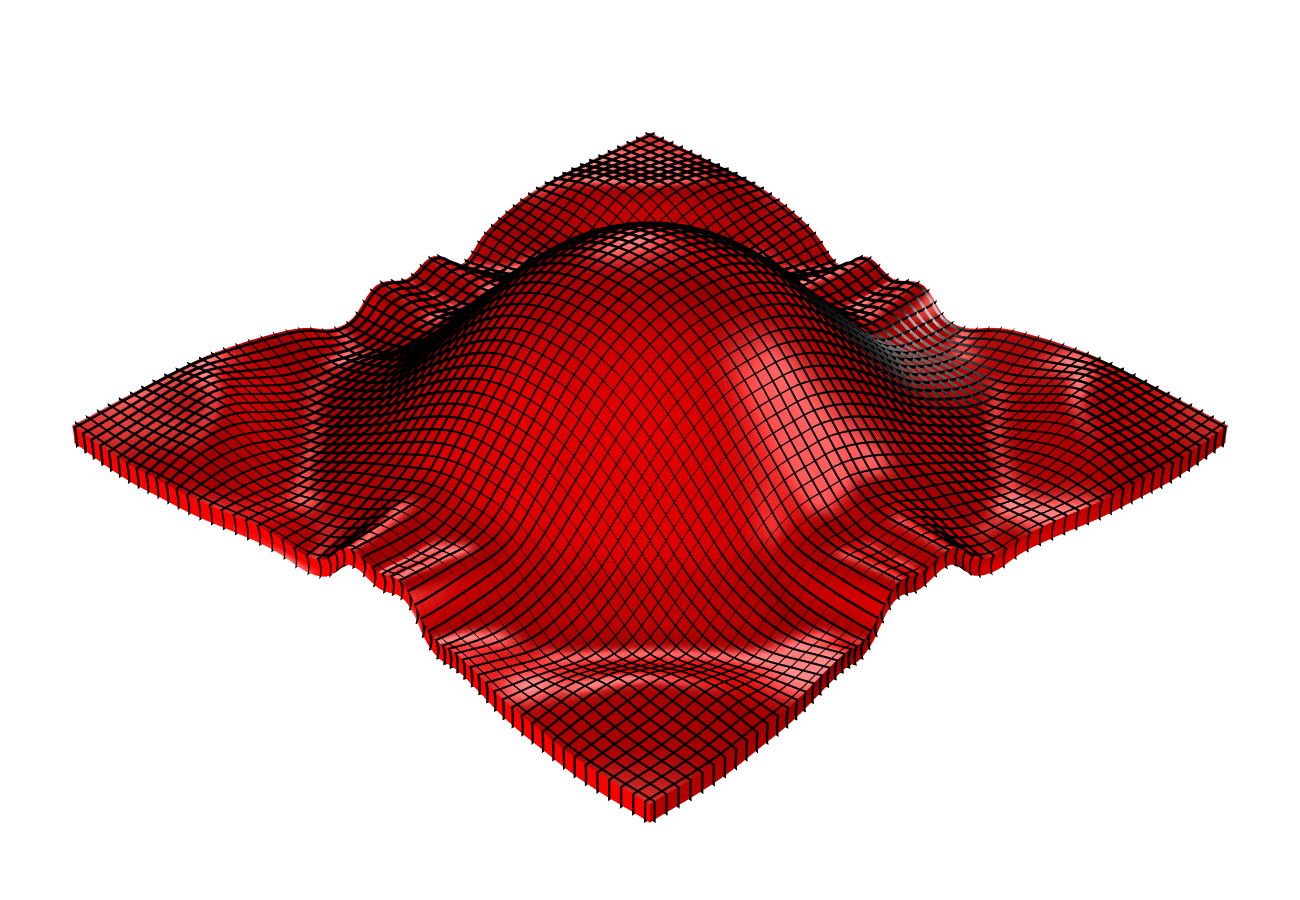}  & \includegraphics[trim=0 30 0 80,width=4.8cm]{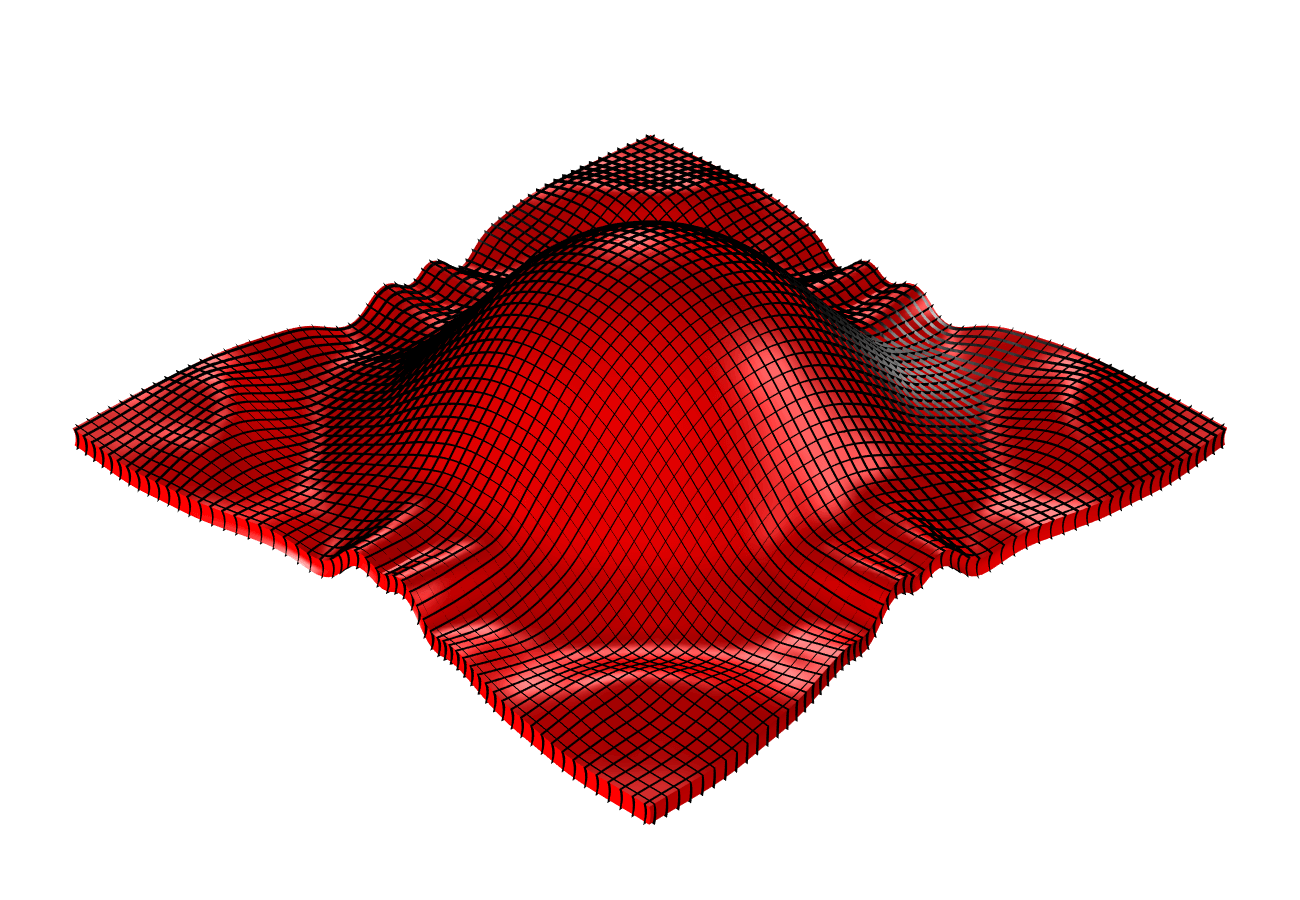} & \includegraphics[trim=0 30 0 80,width=4.8cm]{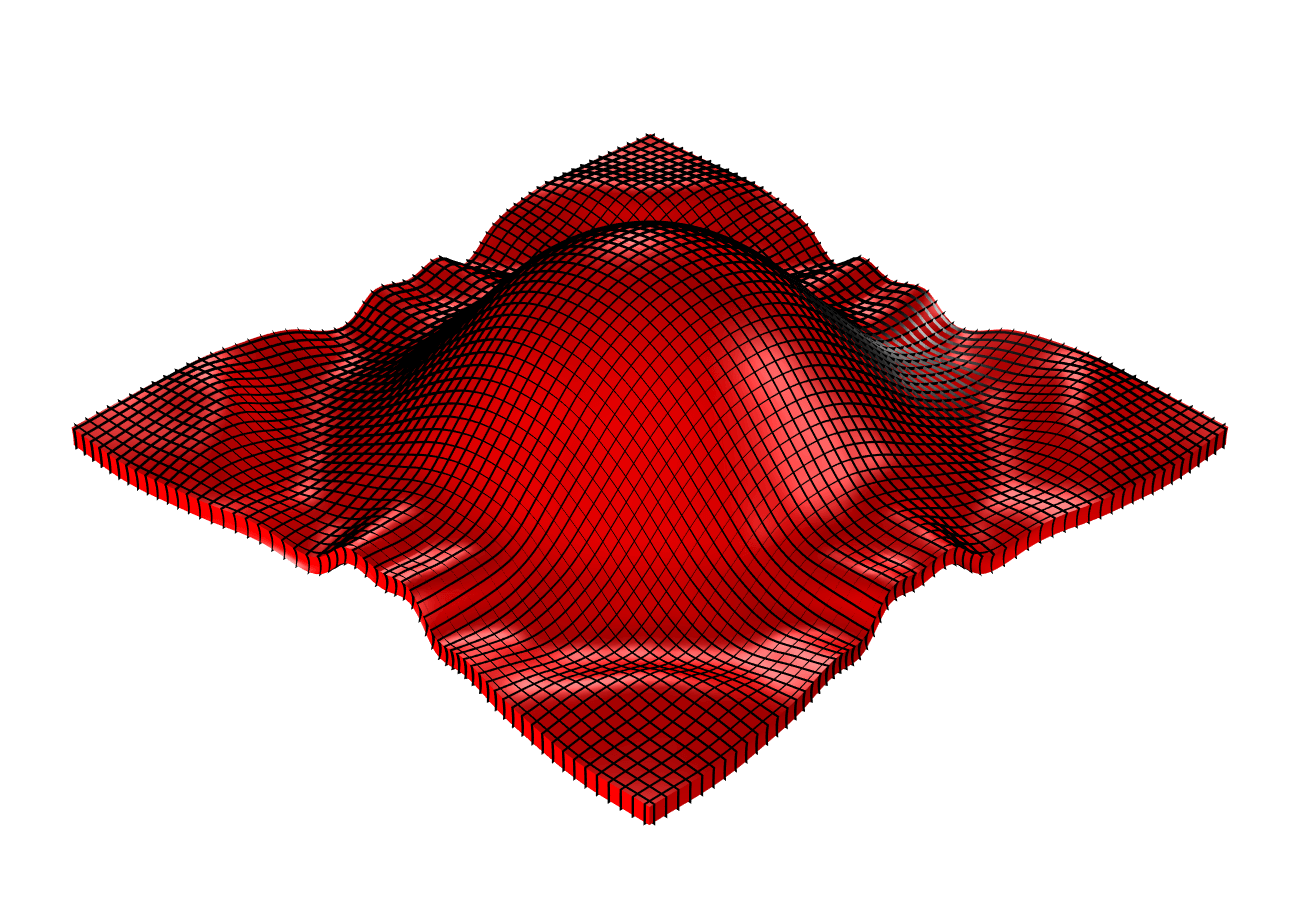} \\\hline
		\end{tabular}
		\par\end{centering}
	
	\caption{ Mesh-dependency of the first and the second gradient model with shape functions with augmented continuity.\label{augm}}	
\end{figure}

In this subsection, a study of first and second gradient solutions obtained considering \textit{augmented continuity shape functions} is presented in Fig.\;\ref{augm}. At first glance, it could seem that the mesh considered is poorer with respect to the previous case, but with the third degree polynomials used as shape function the number of nodes is comparable to the linear case. 

It is possible to see that the first gradient solution still seems to depend on the size of the mesh. The wrinkles are not spikes corresponding to an interface between two mesh-elements, as happened in Fig.$\,$\ref{mapped}, because the \textit{augmented continuity shape functions} impose the smoothness of the strain during the deformation process. 

Fig.\;\ref{augm} explicitly shows that the stability of the model seems to be increased by adding a second gradient energy. Indeed, the wrinkling phenomenon is controlled by the second gradient terms and the corresponding result appears to be mesh-independent even  with a small constitutive parameter ($\alpha=0.1\,N$). The fact that second gradient terms stabilize the numerical onset of wrinkling so producing more realistic results is not surprising. Indeed, the presence of an out-of-plane bending stiffness of the yarns (which is of course evident from a phenomenological point of view) makes energetically expensive the formation of wrinkles. On the other hand, since no energetic cost is associated to out-of-plane bending withing first gradient theories, the onset of a myriad of wrinkles is allowed  even if this solution deviates from experimental evidence. If the value of $\alpha$ is increased, the results obtained with the different meshes considered are the same as in Fig.$\,$\ref{First} and, therefore, they are not included here.

The presented results are very promising but, it must be noted that the \textit{augmented continuity shape functions} are just a workaround for the real problem which is that of implementing robust finite elements for the simulation of woven composite reinforcements in view of structure design. The continuity of the derivatives is weakly imposed and it is, therefore, not strictly granted. A study on the validity of a model implementing the \textit{augmented continuity shape functions} should be made, even if it seems that the presented model is reliable in the description of the analyzed phenomenon. Alternative methods to stabilize the solution can also be found in the literature \cite{mathieu2016stability} consisting in the insertion of small structural elements (such as beams) in the interior of the FE in the direction of the yarns, so indirectly accounting for their bending stiffness.
 
\subsubsection{Influence of cutting the corners on the onset of wrinkling for first and second gradient solutions}

\begin{figure}[H]
	\begin{centering}
		\begin{tabular}{|c|c|c|c|}
							\hline 		&&&\\ \rotatebox{90}{\hspace{1cm}Mesh}&
							\includegraphics[trim=0 60 0 60,width=4.8cm]{Immagini/Deformed1GCont2Mesh.png}  & \includegraphics[trim=0 60 0 60,width=4.8cm]{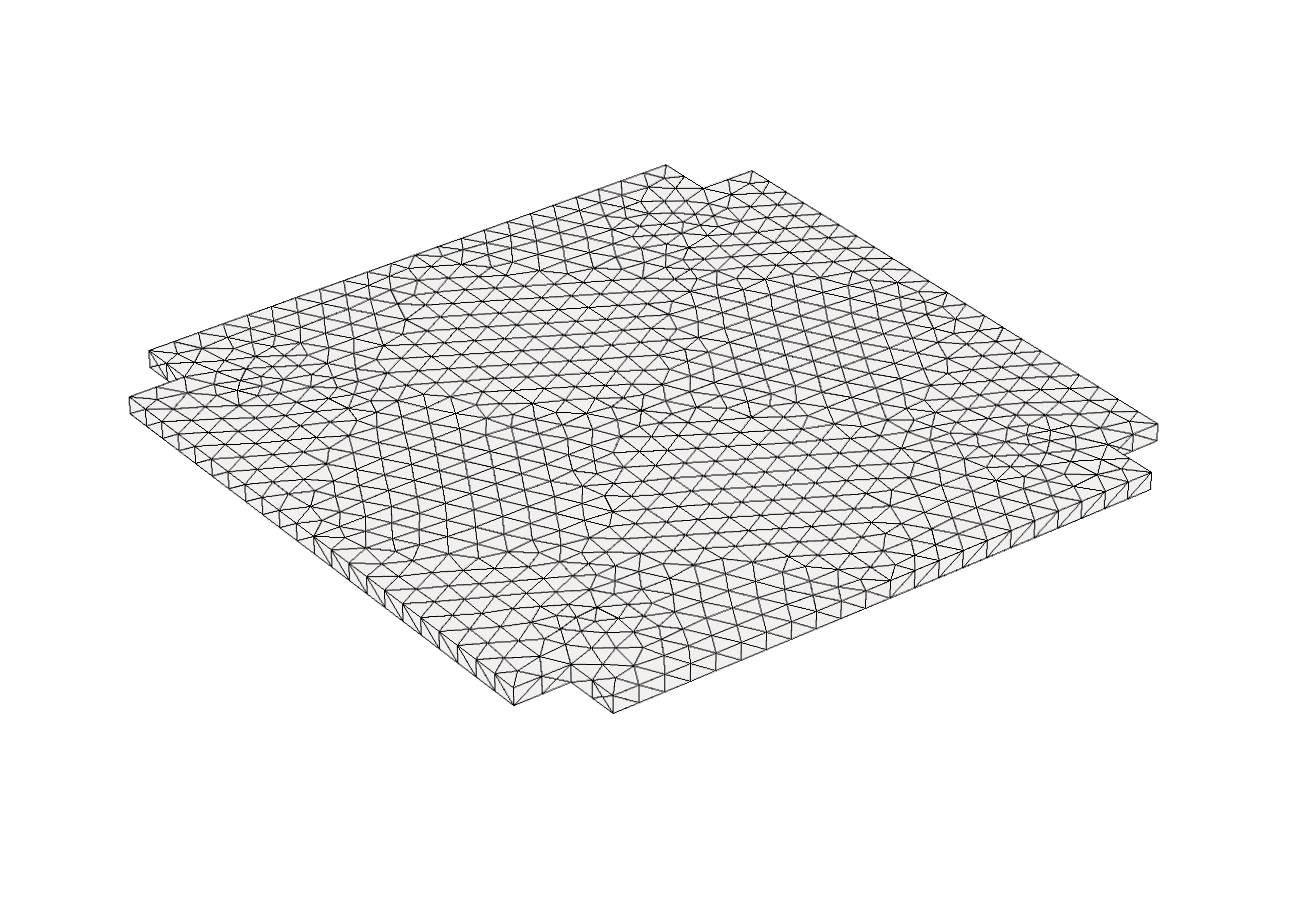} & \includegraphics[trim=0 60 0 60,width=4.8cm]{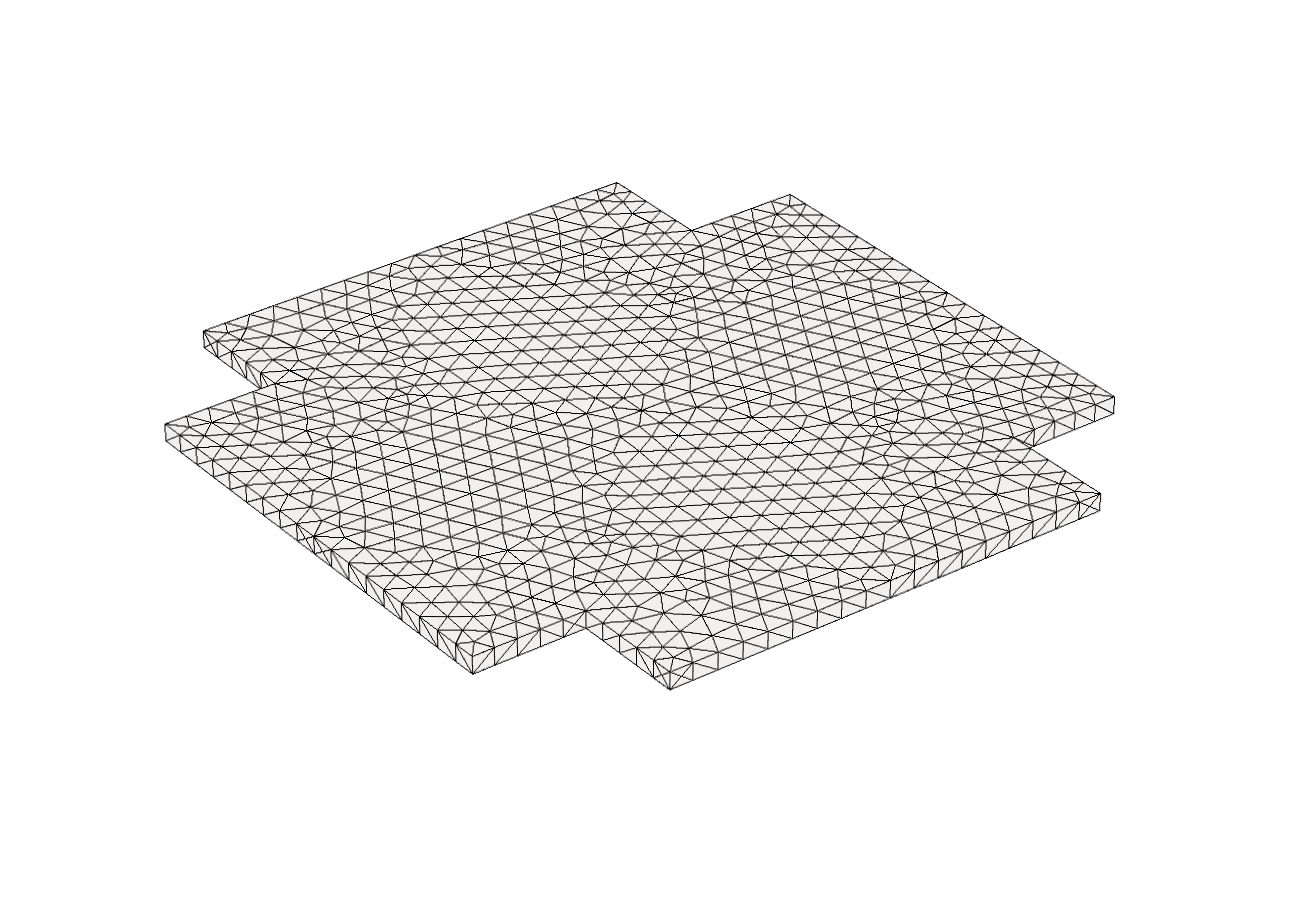} \\\hline&&&\\\rotatebox{90}{\hspace{0.5cm}$1^{st}$ gradient}&
			\includegraphics[trim=0 60 0 60,width=4.8cm]{Immagini/Deformed1G.png}  & \includegraphics[trim=0 60 0 60,width=4.8cm]{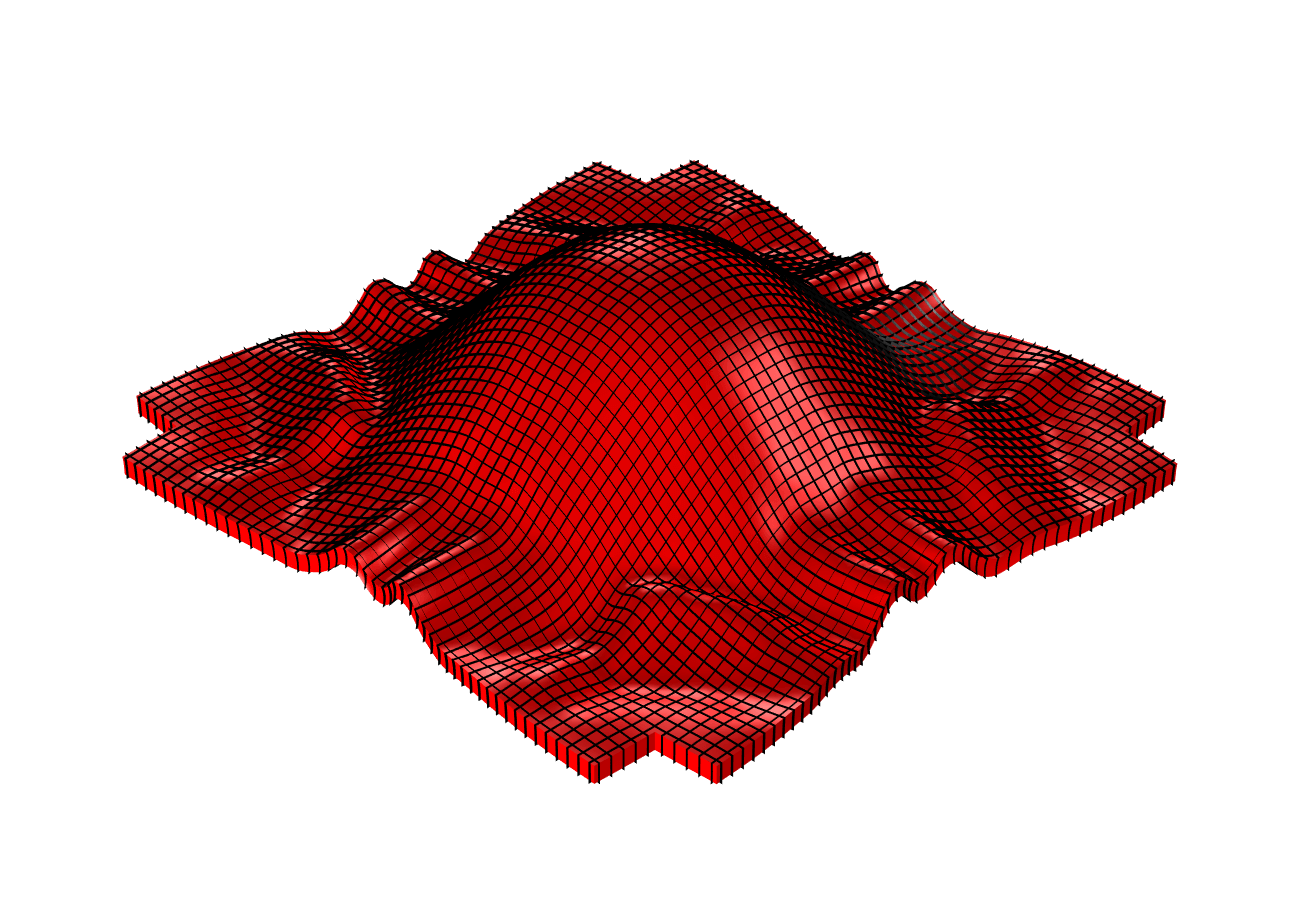} & \includegraphics[trim=0 60 0 60,width=4.8cm]{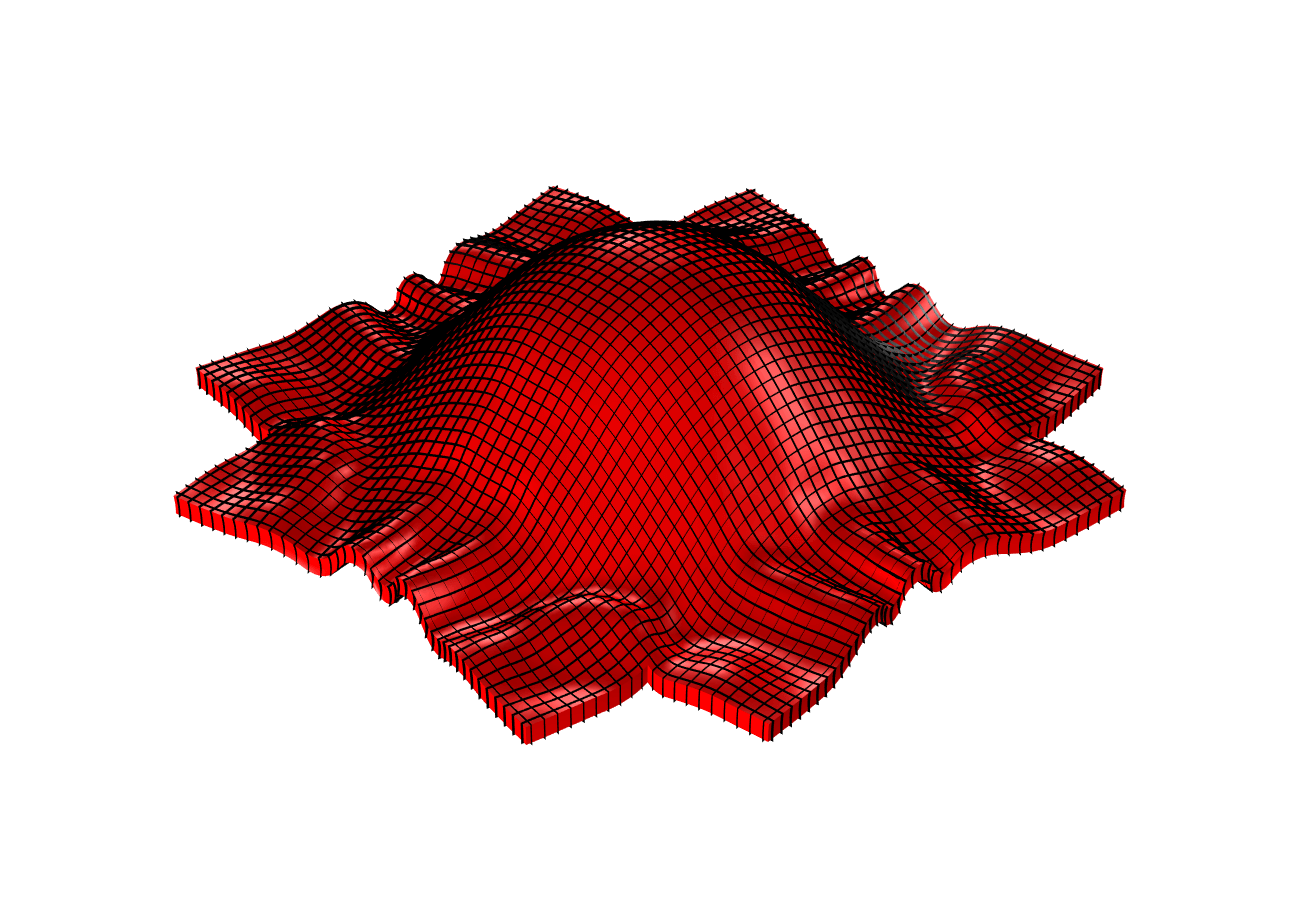}  \\\hline&&&
			\\\rotatebox{90}{ $2^{nd}$ gradient $\alpha=1\,N$}& 	\includegraphics[trim=0 30 0 80,width=4.8cm]{Immagini/Deformed2G1N.png}  & \includegraphics[trim=0 30 0 80,width=4.8cm]{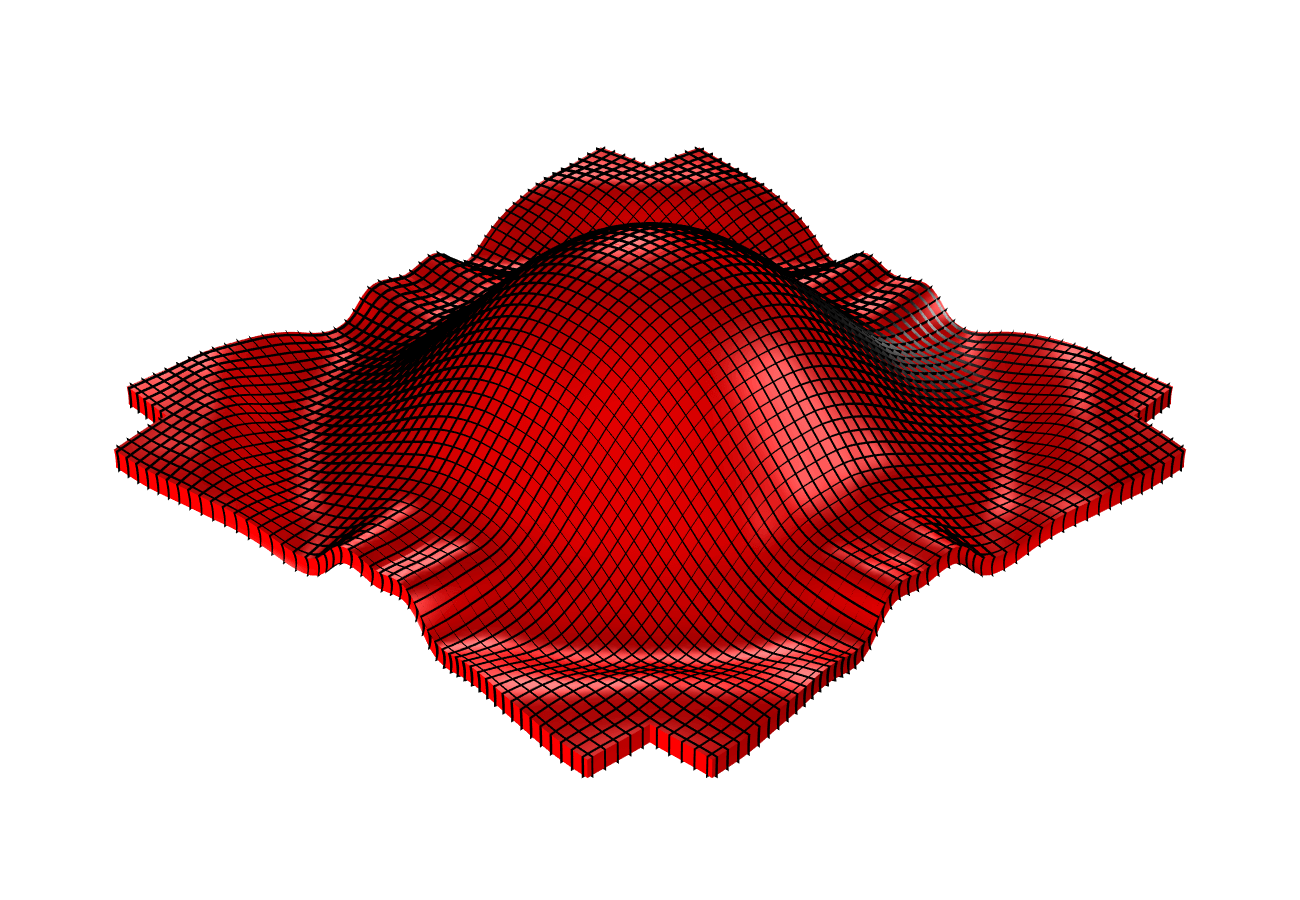} & \includegraphics[trim=0 30 0 80,width=4.8cm]{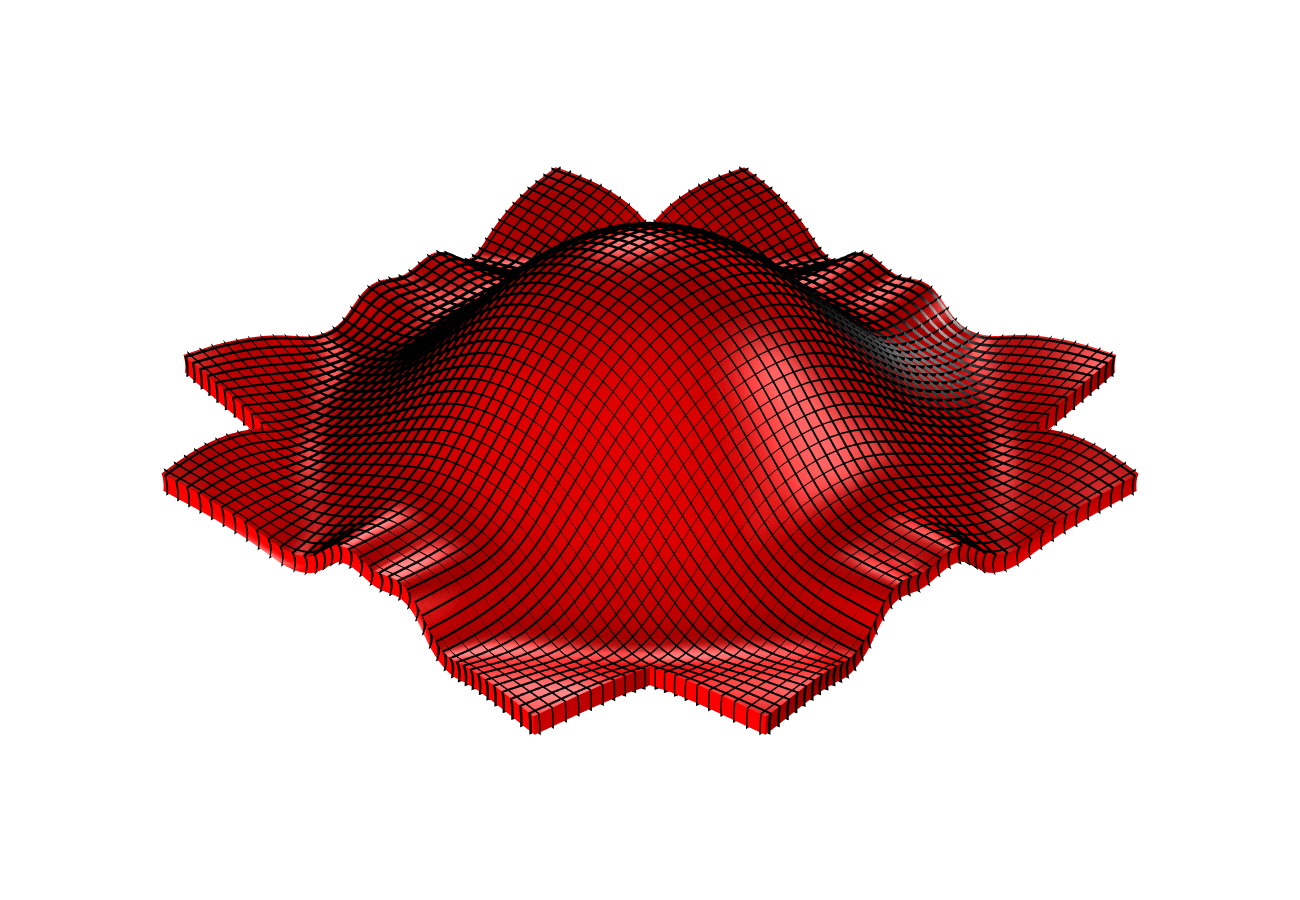} \\\hline
		\end{tabular}
		\par\end{centering}
	
	\caption{ Influence of cutting the corners on the onset of wrinkling for the first and the second gradient model with shape functions with augmented continuity.\label{corners}}	
\end{figure}

During experimental testing, it is a spread routine to cut the corners of the specimen as shown in the first line of figure \ref{corners}, see \cite{charmetant2012hyperelastic}. This change in the geometry can have an influence on the onset of wrinkling during the deep drawing of the fabric. It is possible to notice that cutting the corner leads to a slightly reduced amount of wrinkling in the first gradient model, while for the second gradient model the wrinkling is already not relevant and therefore almost no influence is seen by the cutting of the corners. The considerations concerning the dependence of the solution from the size of the mesh remain the same as in the previous subsections both for the first and second gradient case.

\section{Conclusions}

The woven fabrics posses  a huge potential due to their very specific characteristics. However, there is still no common agreement concerning the approach to be adopted for the modeling of such materials during their dry preforming. The classical models used for the woven composite preforming fail to describe a wealth of observed experimental evidences. In particular, the description of the wrinkling phenomenon is one of the weakest points of such models.

The introduction of terms in the energy depending of higher order derivatives are useful to describe the deformation energy associated to the micro-structure. In the authors' opinion, this possibility is of fundamental importance for an accurate description of the preforming of the woven fabrics as well as for the control of the wrinkling phenomenon. In this paper, the results obtained with the introduction of an energy depending on the in-plane and out-of-plane curvature of the yarns are presented and it is shown how the onset of the wrinkling during the deep drawing can be controlled so as to reproduce real experimental evidence. 

In the literature, second gradient models have already been used to describe specific experimental behaviors of fibrous composite reinforcements and this article provides additional evidences supporting the potential of the use of this type of models for such materials. It is the authors' belief that it should be considered a proven fact that the addition of simple second gradient energy terms could be critical to describe numerous observed phenomena such as the one analyzed here. However, the simple quadratic energy hereby used is not likely to be completely sufficient for the general description of the non-linear behavior of the fabrics, but it is descriptive enough to understand the usefulness of a second gradient model for such materials. Further studies are needed to find more complex non-linear second gradient constitutive laws fully descriptive of the mechanical behavior of the woven fabrics at very large strains and subjected to different boundary and/or loading conditions.

\footnotesize

\let\stdsection\section
\def\section*#1{\stdsection{#1}}

\bibliography{library}

\bibliographystyle{plain}

\let\section\stdsection

\end{document}